\documentclass[aps,prd,nofootinbib,reprint,superscriptaddress,preprintnumbers, onecolumn]{revtex4-2}
\pdfoutput=1
\usepackage[utf8]{inputenc}
\usepackage{amsmath,amssymb}
\usepackage{epsfig}
\usepackage{comment}
\usepackage{graphicx}
\usepackage[colorlinks,citecolor=blue]{hyperref}
\usepackage{xcolor}
\usepackage{subfigure}
\usepackage{cleveref}
\usepackage{url}
\usepackage{dcolumn}
\usepackage{bm}
\usepackage{soul}
\usepackage{slashed}
\usepackage{siunitx}
\usepackage[normalem]{ulem}
\usepackage{mathtools}

\newcommand{\ketL }[1]{| #1\, -  \,\rangle}
\newcommand{\ketR }[1]{| #1\, + \, \rangle}

\newcommand{\ketRL}[1]{|#1 \pm \rangle}
    
\newcommand{\braL }[1]{\langle #1+|}
\newcommand{\braR }[1]{\langle #1- |}

\newcommand{\braRL}[1]{\langle  #1 \mp|}

\newcommand{\braketRR}[2]{
      \langle#1-|#2 +\rangle
    }

\newcommand{\braketLL}[2]{
      \langle#1+|#2-\rangle 
    }

\newcommand{\braketpm}[2]{
      \langle#1\pm|#2\pm\rangle 
    }

\DeclareFontFamily{U}{mathx}{}
\DeclareFontShape{U}{mathx}{m}{n}{<-> mathx10}{}
\DeclareSymbolFont{mathx}{U}{mathx}{m}{n}
\DeclareMathAccent{\widehat}{0}{mathx}{"70}
\DeclareMathAccent{\widecheck}{0}{mathx}{"71}

\begin{document}

\preprint{FERMILAB-PUB-25-0406-T}

\title{Exploring Quantum Statistics for Dirac and Majorana Neutrinos \\using Spinor-Helicity techniques}

\author{Innes Bigaran}
\email{ibigaran@fnal.gov}
\affiliation{Theoretical Physics Department, Fermilab, Batavia, IL, USA}
 \affiliation{Department of Physics \& Astronomy, Northwestern University, Evanston, IL, USA}
 
\author{Stephen J. Parke}
\email{parke@fnal.gov}
\affiliation{Theoretical Physics Department, Fermilab, Batavia, IL, USA}

\author{Pedro Pasquini}
\email{pasquini@ifi.unicamp.br}
\affiliation{ Instituto de F\'isica Gleb Wataghin - Universidade Estadual de Campinas (UNICAMP), 13083-859, Campinas SP, Brazil}

\begin{abstract}

Recently, there has been interest in the applicability of quantum statistics to distinguish Dirac from Majorana neutrinos in multi-neutrino final states. In particular, debate has arisen over the validity of the Dirac-Majorana confusion theorem in these processes, \emph{i.e.}, that any distinction between the Dirac and Majorana processes goes to zero as the neutrino mass goes to zero. Here we approach this problem equipped with spinor-helicity methods generalized for massive Dirac and Majorana fermions. We explicitly calculate all helicity amplitudes, and their squares, for the decay of a light scalar particle to two neutrinos and two oppositely charged leptons. This allows us to pinpoint the crucial steps which could lead to claims of a violation of the confusion theorem. We show that if the correct anti-symmetrization of Dirac to Majorana amplitudes is used, identification of which is clear in this framework, and all relevant contributions are appropriately summed, a scalar decay into two charged leptons and two neutrinos satisfies the Dirac-Majorana confusion theorem.  
\end{abstract}

\maketitle
\section{Introduction}

The nature of the neutrino as Dirac or Majorana is one of the most important open questions in neutrino physics. Are the neutrinos Dirac fermions with conservation of lepton number, or Majorana fermions
with lepton number violated? 
Answering this question is extremely challenging since any known difference between Dirac and Majorana neutrinos in a physical observable, within the Standard Model~(SM) framework, is suppressed by the neutrino mass scale -- which is smaller than one millionth of the mass of the electron. 
This is the Dirac-Majorana confusion theorem (DMCT), first presented by Kayser and Shrock in ref.~\cite{Kayser:1982br} and ref.~\cite{Shrock:1982sc}, respectively. For this theorem, there is not presently a general,
model-independent and process-independent proof.\\   

The quest to shed light on the nature of the neutrino motivates searching for processes that could violate the DMCT, and make distinguishing Dirac from Majorana more experimentally accessible. For processes which violate lepton number, and are zero otherwise, consistency with the DMCT follows from the neutrino mass as a spurion of symmetry breaking -- for example, neutrinoless double beta decay~\cite{Schechter:1981bd,Nieves:1984sn,Takasugi:1984xr, kayser1987proc,Rosen:1992qa}. However, in processes involving multiple final-state neutrinos as well as addition observable particles, it has been proposed that differences in the quantum statistics of Dirac and Majorana neutrinos could give rise to correlations of the observable particles that are not subject to this suppression. For example, in a recent sequence of papers~\cite{Kim:2021dyj,Kim:2023iwz, Kim:2024tsm} the authors claim to have discovered a violation of the DMCT in $B_0$ decays to two charged leptons and two neutrinos. This claim has been disputed in ref.~\cite{Akhmedov:2024qpr}, giving general arguments that the above violation of the DMCT is not physical.  In this work, we revisit the impact of quantum statistics on decays to two-neutrino final states together with additional charged leptons using spinor helicity techniques. \\

Notably, the DMCT applies to interactions within the framework of the SM, where left-chiral neutrinos interact only via the electroweak force, and right-chiral neutrinos are sterile. The chiral nature of the SM electroweak interactions make spinor-helicity techniques particularly suited to this application. In this work, we use massive spinor helicity formalism, in which massive spinors can be constructed as a sum of massless spinors with shifted momenta -- a framework introduced by Mahlon and Parke in ref.~\cite{Mahlon:1995zn} for calculating top-quark angular correlations. For massless spinors, helicity and chirality are equivalent, and thus tree-level calculations of processes involving chiral interactions are notably simplified. While the greatest gain in computational efficiency arises for processes involving chiral interactions, the methodology introduced here for neutrino final-state calculations is not limited to such cases, and can be readily extended to frameworks beyond the Standard Model (BSM). \\

In this work, we primarily highlight a demonstrative example of decays into multi-neutrino final states. This is the decay of a neutral scalar particle, with a mass roughly that of the $B_0$ meson, into two charged leptons and two neutrinos via off-shell $W$ bosons. We consider the cases of Dirac and Majorana neutrinos, with particular emphasis on the phases of the spinors so as to calculate the Majorana amplitudes from Dirac amplitudes with correct, physical, anti-symmetrization. We explicitly give the Dirac and Majorana amplitudes for each spin configuration, and their square. We show that mathematically there are two ways to construct this anti-symmetrization: one that leads to a set of amplitudes that make consistent physics sense, and the other leads to a set of amplitudes with pathologies that make them unphysical. Elucidating the physical distinction between these two choices is particular clear using the spinor helicity formalism evaluated in the helicity basis.\\

The outline of this paper is as follows. In Section~\ref{sec:Spinors}, we give a brief summary of the spinor-helicity method for massive fermions, before proceeding to Section~\ref{sec:Scalar} where we discuss light neutral scalar decay.
In Section~\ref{sec:Dir} we give the four independent amplitudes for our scalar particle decay assuming the neutrinos are Dirac fermions, their square and sum over the spins, assuming the charged leptons are massless for simplicity. In Section~\ref{sec:AntiS}, anti-symmetrization possibilities of the Dirac amplitudes to form Majorana amplitudes are discussed, and the physical anti-symmetrization is used in Section~\ref{sec:Maj} to construct a set of three independent amplitudes. 
The phenomenology and the difference between Dirac and Majorana neutrinos is discussed in Section~\ref{sec:Pheno}, concluding that the DMCT is not violated in these decays. In Section~\ref{sec:CLMass} we show that considering charged leptons with non-zero mass does not change the conclusions above.
In Section~\ref{sec:b0decay} we explicitly apply the results above to the case of $B_0$ leptonic decays, and in Section~\ref{sec:BSM} briefly discuss how BSM physics can change this picture. We conclude in Section~\ref{sec:Concl}, and also include a selection of appendices that highlight more detailed aspects of our methods, arguments, and confirmation of techniques.


\section{Spinor-Helicity Technics for Massive Dirac or Majorana Fermions}
\label{sec:Spinors}

We follow the spinor-helicity techinique summarized by Mangano and Parke in ref.~\cite{Mangano:1990by}, which we briefly review here. Let $u(q)$ and $v(q)$ represent spinors for a massless fermion and anti-fermion, respectively, each with four-momentum $q$ ~($q^2=0$). The complete set of massless spinors is defined as
\begin{align}
    \ketRL{q}\equiv \gamma_{R/L} u(q)= \gamma_{R/L}v(q),  \quad \text{and}
    \quad \braRL{q}\equiv \bar{u}(q)\gamma_{R/L}=\bar{v}(q)\gamma_{R/L},\label{Eq:helicitystates}
\end{align}
where $\gamma_{R/L} \equiv  \frac{1}{2} (1\pm \gamma_5)$, \emph{i.e.} the $\pm$ notation denotes the chirality of the state, which is equivalent to helicity only in the massless limit. Note that, for example, $\ketL{q}$ represents both a negative helicity massless fermion and a positive helicity massless anti-fermion. From these definitions it follows that 
\begin{align}
   \gamma_R \ketL{q}= 0, &\quad \gamma_L\ketR{q}= 0,  \quad \braR{q}\gamma_L=0\quad \braL{q}\gamma_R=0\, 
\quad {\rm and} \quad 
  \braketRR{q}{q}=0,
\notag  \\[2mm]
\text{and} \quad   \slashed{q} &= \ketR{q}\braL{q}+ \ketL{q}\braR{q} \,.
\end{align}
Considering also an additional massless momentum $k$, where $k^{2}=0$, then
\begin{align}  
\braketpm{k}{q} &= 0
,\quad 
\braketLL{k}{q}= - \braketLL{q}{k}
,\quad 
  \braketRR{k}{q}=(\braketLL{q}{k})^*
\quad \text{and} \quad 
\braketRR{k}{q}\braketLL{q}{k} = 2 (k\cdot q),
\end{align}
thus $\braketRR{k}{q}$ and $\braketLL{q}{k}$ are complex conjugate square roots of $(2 k \cdot q)$. 
Many more identities can be derived for these spinors, see for example ref.~\cite{Mangano:1990by} where a more compact notation was used with  $\braketRR{k}{q} = \langle k q \rangle$ and
$ \braketLL{k}{q} = [k q ] $. Another alternative notation sometimes used is that $\braketRR{k}{q}  = k_\alpha\, q^\alpha$ and $\braketLL{k}{q} = \tilde{k}^{\dot{\alpha}}\, \tilde{q}_{\dot{\alpha}}$, see for example ref.~\cite{Dreiner:2008tw}.
 Here we use a more explicit notation to emphasize the simplicity and understandability of our results to a wide readership.\\

For a massive fermion, the spinors are constructed by decomposing a massive momentum 4-vector, $p$, with $p^2=m^2$, into two massless momentum 4-vectors.
One of these momenta could be a rescaled momentum of a massless particle for the process, as was done by Kleiss and Stirling in ref.~\cite{Kleiss:1985yh}. 
Here we adopt a different choice: constructing the two massless momentum using the spin 4-vector for the massive fermion, $s$, as was done by Mahlon and Parke in ref.~\cite{Mahlon:1995zn}. We use $\widehat{p}$ and $\widecheck{p}$ to denote these massless momenta and define them as
\begin{align}
    \widehat{p}\equiv \frac{1}{2}(p+ms), \quad \widecheck{p}\equiv \frac{1}{2}(p-ms)
\end{align}
which are massless momenta, $\widehat{p}^{\,2}=0, ~\widecheck{p}^{\,2}=0$, as $p\cdot s =0$ and $s^2 =-1$. Also, $2\,\widecheck{p} \cdot \widehat p =  m^2$ and
\begin{align}
  p = \widehat{p} + \widecheck{p}, 
\quad \text{and} \quad
  s = (\widehat{p} - \widecheck{p})/m .
\end{align}
Thus we have traded the two non-null four vectors, $(p,s)$, for two massless four vectors
$(\widehat{p}, \widecheck{p})$, this allows us to use the full powerful machinery of the spinor-helicity method.\\

With these two momenta, one can decompose all massive spinors into two chiral components with each being a spinor for one or other of these  massless momenta, $(\widehat{p}, ~\widecheck{p})$. There is an important phase multiplying each of the massless spinors as follows:

\begin{align}
  \overline U_\uparrow(p, s) 
  &=~ \, e^{-i\psi} \,\braL{\widehat{p}} + e^{i\psi}  \braR{\widecheck{p}} 
  \,, \quad\;
 V_\uparrow (p, s)=~\, e^{-i\psi} \ketL{\widehat{p}} +   e^{i\psi}  \ketR{\widecheck{p}} 
  \,,
 \notag \\[3mm] 
\overline U_\downarrow(p,  s)
  & = - e^{- i\psi } \braL{\widecheck{p}} + e^{i\psi} \braR{ \widehat{p}}  
     \,,  
  \quad
  V_\downarrow (p, s)= -e^{-i\psi}  \ketL{\widecheck{p}} + e^{i\psi}  \ketR{\widehat{p}}  
   \, ,
  \label{eq:spinors}
  \end{align}
  where the phase $\psi$, which is momentum and spin axis dependent,  is given by 
  $$ \psi \equiv \frac1{2}\text{Arg}[ \braketLL{\widehat{p}}{\widecheck{p}}/m ] \, . $$
 This way of writing the spinors is particularly powerful if the interactions are of a chiral nature as one of the two terms in the spinors is annihilated.

Further discussion on the spinor conventions and the square root choices in the definition of $\psi$ are given in  Appendix~\ref{app:morespinors}. 
For calculational purposes, it is convenient to factor out an $\exp[-i\psi]$ for all spinors since they will become an overall phase,
$\exp[-i(\psi_1+\psi_2)]$, for all Dirac and Majorana amplitudes.
\\

The above applies for a general spin axis. In the main text of this paper, we will use the language of the helicity basis because it reveals a smooth limit for when the mass goes to zero. However our results are valid for a general basis.  For the helicity basis, the spin axis in the rest frame of the particle is in the boost direction, $\tilde{n}_p$, such that $s=(0,\tilde{n}_p)$,  with $s^2=-1$. Therefore after applying this boost
we have
  \begin{align}
  p 
&=
  \gamma m\, (1, \beta \tilde{n}_p),   \quad s=\gamma (\beta, \tilde{n}_p) \quad \text{giving} 
\notag  \\
  \widehat{p} 
&\equiv 
  \frac1{2} \gamma (1+\beta)m\,(1, +\tilde{n}_p) \quad  \xRightarrow{\gamma\to \infty} \quad \quad \approx \gamma m(1, \tilde{n}_p)  =E(1,\tilde{n}_p), 
\label{eq:helbasis}\\
  \widecheck{p} 
\label{eq:p_limits}
&\equiv 
  \frac1{2} \gamma (1-\beta)m\,(1, -\tilde{n}_p) \quad \xRightarrow{\gamma\to \infty} \quad \quad \approx \frac1{4 \gamma} m(1, -\tilde{n}_p)= \left(\frac{m}{2E}\right)^2 E (1,-\tilde{n}_p), \notag
\end{align} 
with  $p= \widehat{p} +\widecheck{p}$ as required. For the helicity basis, spin up ($\uparrow$) will be denoted by an $R$, and spin down ($\downarrow$) will be denoted by an $L$. The expressions on the right give the leading terms in the ultra-relativistic limit, $\gamma \rightarrow \infty $. In this limit, the momentum $\widehat{p}$ points in same direction with very nearly the same energy as the massive particle, whereas 
$\widecheck{p}$ points in the opposite direction with an energy suppressed by factor of  $(m/2E)^2$. By ignoring the  $\widecheck{p}$ terms one recovers the massless limit.

\section{Light Scalar Decay as an Instructive Example}
\label{sec:Scalar}

Consider a light scalar particle, $\Phi$, that decays into two off-shell $W$-bosons and then each $W$ decays into a charged lepton and a neutrino, so that the final state is two charged leptons and two neutrinos, as follows: 
\begin{align}
\Phi(Q) \rightarrow \{W^+\}^* \{W^-\}^* \rightarrow  l^+(\bar{k}) + \nu_1(p_1)  + l^-(k)  + \nu_2(p_2) 
\end{align}
with $M_\Phi \ll M_W$. The four momenta of the two massless charged leptons are $k$ and $\bar{k}$, for the neutrinos $p_1$ and $p_2$ and $Q$ for the scalar particle.  The two neutrinos, $\nu_1$ and $\nu_2$, have the same flavor and mass $m_\nu$.
For Dirac neutrinos we might call $\nu_1$ or $\nu_2$ a neutrino and the other an anti-neutrino, whereas for Majorana neutrinos this particle-antiparticle designation is not available, that is why we use $\nu_1$ and $\nu_2$. The $\Phi$ particle can be considered a light Higgs-like scalar and the coupling of the $W$'s to the lepton-neutrino pair is left-handed as in the Standard Model. The mass of the $\Phi$ particle is chosen well below the W-boson mass so that one cannot kinematically pair either of the neutrinos with one of the charged leptons as both possibilities are kinematically equally favored. For simplicity we initially set the charge lepton masses to zero, so that we can purely concentrate on the effects of the neutrino mass and the distinction between Dirac and Majorana neutrinos. Later we will address how the charge lepton masses effect this process and also discuss allowing for the possibility of the effects of adding a right-handed $W^\prime$ or  $Z^\prime$ boson.\\

\subsection{Massive Dirac Neutrinos}
\label{sec:Dir}

For Dirac neutrinos we have designated $\nu_1$ as the neutrino and $\nu_2$ as the anti-neutrino, therefore the amplitude for this decay,  ${\cal D}[(p_1,s_2), \overline{(p_2,s_2)}] $, (the bar over the $(p,s)$ indicates it is an  anti-fermion) is given by
\begin{align}
 {\cal D}[(p_1,s_2), \overline{(p_2,s_2)}] \sim&  ~ \overline{u}(k) \, \gamma^\mu_L \,V(p_2, s_2) ~~.~~ \overline{U}(p_1,s_1)\, \gamma^\mu_L \,v(\bar{k}) \notag \\
  = & ~ \braR{k} \gamma^\mu_L \,V(p_2, s_2) ~~ . ~~ \overline{U}(p_1,s_1)~ \gamma^\mu_L  \ketL{~\bar{k}}  \label{eq:DiracLine2}\\
  =& - 2 \braL{~\bar{k}}V(p_2, s_2) ~~. ~~  \overline{U}(p_1,s_1) \ketR{k} 
  \,,
  \label{eq:Fierz}
 \end{align}
where $\gamma_L^\mu = \gamma^\mu \gamma_L$, and a Fierz identity is used in moving from eq.~\eqref{eq:DiracLine2} to \eqref{eq:Fierz}.  We will use the following abreviation for this amplitude: ${\cal D}[S_1, \overline{S}_2]$ where $S_i$ is either $L_i$ or $R_i$ and the momentum is given by $p_i$.  Since the weak interactions are left-chiral, only the left-chiral parts of the spinors in eq.~\eqref{eq:spinors} survive. Notice, these components only contain massless spinors, a situation for which the spinor-helicity techniques developed for massless quarks and gluons is ideally suited.   \\

\begin{figure}[t]
\begin{center}
\includegraphics[width=0.95\textwidth]{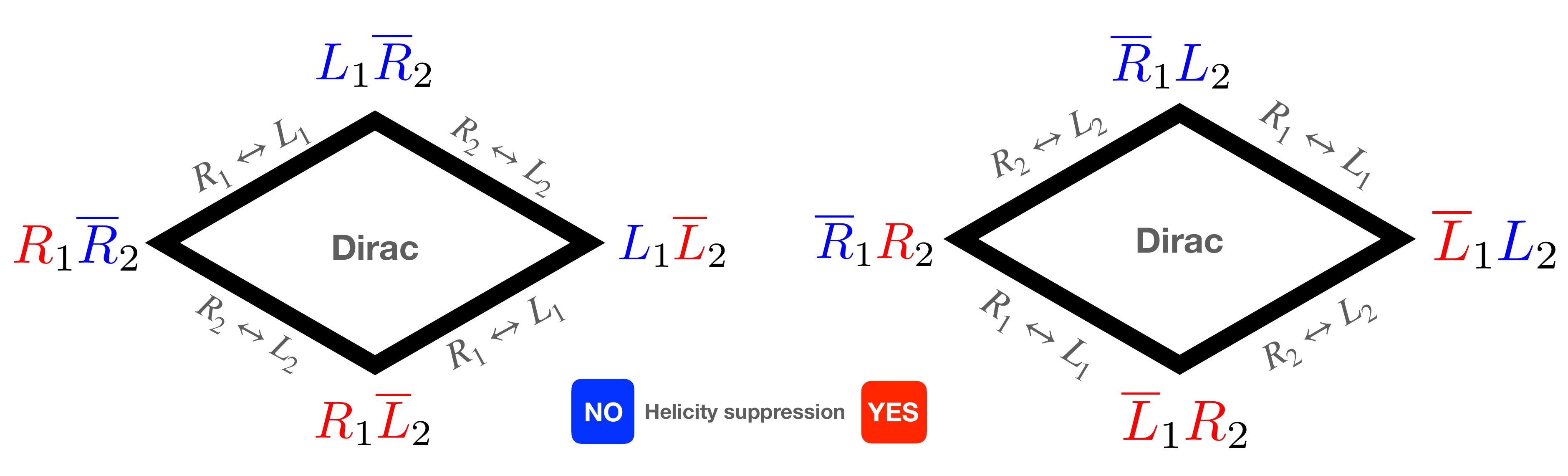}
\caption{The eight Dirac amplitudes, obtained from four independent functions. Left is when $\nu_1$ is the fermion and $\nu_2$ is the anti-fermion and right is vice versa. 
The letter represents the helicity ($L$ or $R$), a bar is added if its an anti-fermion. The color coding red (blue) denotes when there is (no) helicity suppression. One moves from one vertex to another by $L \leftrightarrow R$ for one or other of the neutrinos, as indicated. Note, each set of four amplitudes is closed and complete and the right set can be obtained from the left set by interchanging $\nu_1$ and $\nu_2$. For given phase space point, $(p_1,p_2)$, both must be included unless you have determined which is a neutrino and which is an anti-neutrino. }
\label{fig:Dir}
\end{center}
\end{figure}

Using the spinors from the previous section, we then have that the Dirac amplitudes in the helicity basis  as in Fig.~\ref{fig:Dir} is given by
\begin{align}
 {\cal D}[\, L_1, \overline{R_2}\, ] 
= & ~~\phantom{-}  2 \braketLL{\bar{k}}{\widehat{p}_2}  \braketRR{\widehat{p}_1}{k} ~ \{\braketLL{\widehat{p}_1}{\widecheck{p}_1} /m_\nu \}  \,, \notag \\
  {\cal D}[\, R_1, \overline{R_2}\, ] 
       = &~~\phantom{-}2  \braketLL{\bar{k}}{\widehat{p}_2} \braketRR{\widecheck{p}_1}{k} ~\{ \braketLL{\widehat{p}_1}{\widecheck{p}_1} /m_\nu \} \,, \notag \\
 {\cal D}[\, L_1, \overline{L_2}\, ] 
= &~ - 2\braketLL{\bar{k}}{\widecheck{p}_2}  \braketRR{\widehat{p}_1}{k} ~\{ \braketLL{\widehat{p}_1}{\widecheck{p}_1} /m_\nu \}  \,,\notag \\
 {\cal D}[\, R_1, \overline{L_2}\, ] 
= &~  \,- 2 \braketLL{\bar{k}}{\widecheck{p}_2} \braketRR{\widecheck{p}_1}{k} ~
\{ \braketLL{\widehat{p}_1}{\widecheck{p}_1} /m_\nu \} \,.
\label{eq:Damps}
\end{align}
\\
A common phase to all of these amplitudes $(-e^{-i(\psi_1+\psi_2)})$ has been removed here.
The factors in the $\{\cdots\}$ are just phase factors which are unimportant for Dirac neutrinos but will be important when we consider Majorana  neutrinos. \\

Squaring these amplitudes gives
\begin{align}
   |\,{\cal D}[\, L_1, \overline{R_2}\, ] \,|^2 & ~\varpropto ~ (k \cdot \widehat{p}_1)(\bar{k}\cdot \widehat{p}_2) \notag \\
   |\,{\cal D}[\, R_1, \overline{R_2}\, ] \,|^2 &~\varpropto ~ (k \cdot \widecheck{p}_1)(\bar{k}\cdot \widehat{p}_2) \notag \\
    |\,{\cal D}[\, L_1, \overline{L_2}\, ] \,|^2  &~\varpropto ~ (k \cdot \widehat{p}_1)(\bar{k}\cdot \widecheck{p}_2) \notag \\
     |\,{\cal D}[\, R_1, \overline{L_2}\, ] \,|^2  &~\varpropto ~ (k \cdot \widecheck{p}_1)(\bar{k}\cdot \widecheck{p}_2)\,,
      \end{align}
where they share a common prefactor. 
Using the expressions for $\widehat{p}$ and $\widecheck{p}$ in the helicity basis given in eq.~\eqref{eq:helbasis}, the relative magnitude of these matrix elements in the ultra-relativistic limit are as follows:
    \begin{align}
 & |\,{\cal D}[\, L_1, \overline{R_2}\, ] \,| :  |\,{\cal D}[\, R_1, \overline{R_2}\, ] \,| : 
 |\,{\cal D}[\, L_1, \overline{L_2}\, ] \,| :  |\,{\cal D}[\, R_1, \overline{L_2}\, ] \,| 
 \notag\\[2mm] & \hspace{2cm}
 \sim \mathcal{O}(1) : \mathcal{O} (1/\gamma_{p_1}) : \mathcal{O}(1/\gamma_{p_2}) : \mathcal{O}(1/(\gamma_{p_1} \gamma_{p_2})),
 \end{align}
  where $\gamma_{p_i}$ is the Lorentz factor for the particle with momentum $p_i$. This demonstrates that there is no helicity suppression in $L\overline{R}$,   single suppression in the  $R\overline{R}$ and$L\overline{L}$ and double suppression in the $R\overline{L}$ amplitudes.\\

   Summing over all helicity states of the neutrinos we have
  \begin{align}
  \sum_\text{spins}  |\,{\cal D}[\,1 \, , \overline{2}\, ]\, |^2 &~\varpropto ~ (k \cdot p_1)(\bar{k} \cdot {p}_2) \,.
   \end{align}
   If we have not determined experimentally that 
   the neutrino with momentum $p_2$ is an anti-neutrino 
   we also need to add the possibility that it is a neutrino, with matrix element squared given by
   \begin{align}
   \sum_\text{spins}  |\,{\cal D}[\,2 \, , \overline{1}\, ]\, |^2 & ~\varpropto ~ (k \cdot p_2)(\bar{k} \cdot {p}_1) \,.
   \end{align}
   So that the full amplitude squared is given by
   \begin{align}
   \sum_\text{spins}  |\,{\cal D}[\,1 \, ,  \overline{2}\, ]\, |^2  +  |\,{\cal D}[\,2 \, , \overline{1}\, ]\, |^2 & ~\varpropto ~
   (k \cdot p_1)(\bar{k} \cdot {p}_2) + (k \cdot p_2)(\bar{k} \cdot {p}_1) \,.
   \label{eq:totD}
 \end{align}
This is the full Dirac amplitude squared for the decay of the $\Phi$ assuming that the neutrinos are Dirac particles, that is, lepton number is conserved. Note that if we integrate this amplitude-squared over the full phase space for $p_1$ and $p_2$, we must divide by two so as to not double count.\\

\subsection{Anti-symmetrization to form Majorana Amplitudes}
\label{sec:AntiS}

\begin{figure}[t!]
\begin{center}
\includegraphics[width=0.45\textwidth]{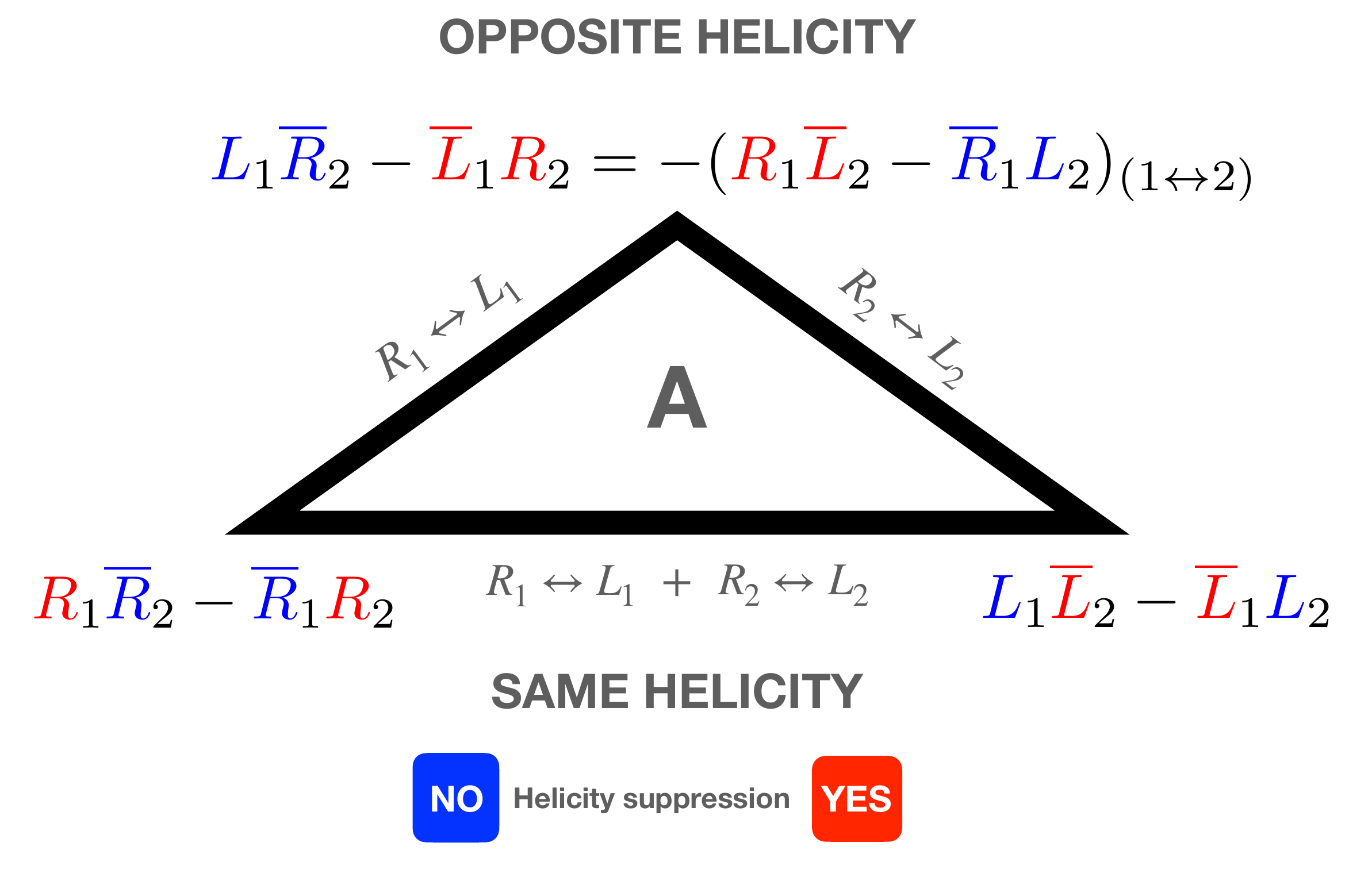} \hspace{1cm}
\includegraphics[width=0.45\textwidth]{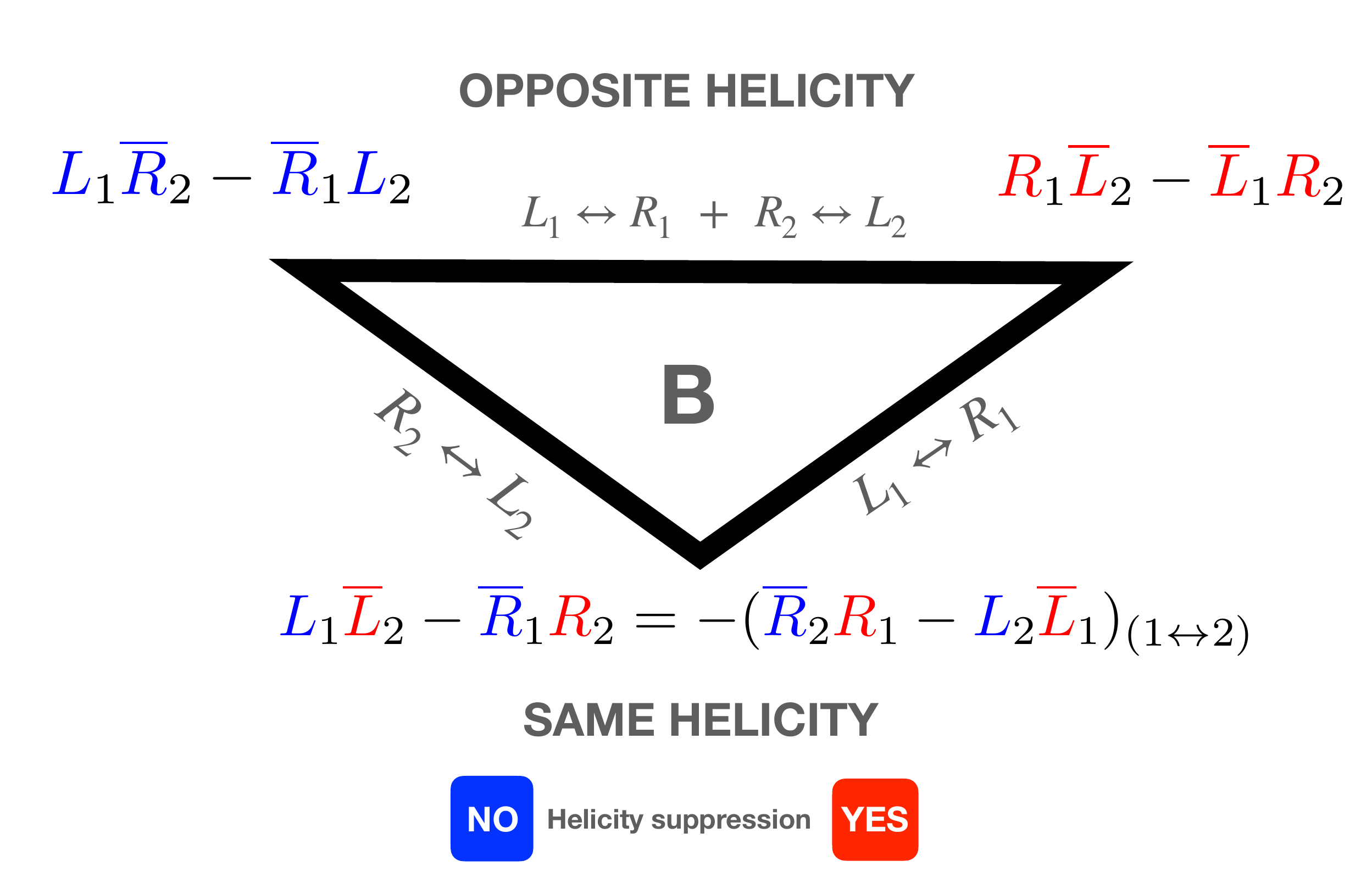}
\caption{ The two sets of anti-symmetrized functions of the Dirac amplitudes with $L_1\overline{R}_2-\overline{L}_1R_2$ (Option A, left) and  $L_1\overline{R}_2-\overline{R}_1L_2$ (Option B, right). From any vertex one can move to another by $L \leftrightarrow R$ for one or both of 1 and 2.  Both sets of functions are closed and complete and all eight Dirac amplitudes are included in each set. As argued in the text, Option A is the correct anti-symmetrization for Majorana neutrinos. }
\label{fig:AntiS}
\end{center}
\end{figure}

We now consider the case that the two neutrinos are Majorana. From the Dirac amplitudes, there are mathematically two ways to anti-symmetrize with the goal to arrive at expressions for the Majorana amplitudes. Beginning with $L_1\overline{R}_2$, one can anti-symmetrize this with either $\overline{L}_1R_2$ or  $\overline{R}_1L_2$. One can then generate the other helicity combinations by exchanging $L \leftrightarrow R$ on one or both of the neutrinos. We illustrate the results  of this exercise in Fig.~\ref{fig:AntiS}:
two sets of three independent functions of the Dirac amplitudes that are both closed and complete.\\

We then ask which one of these sets of functions represent the amplitudes for a process with two  Majorana neutrinos.  For such processes there are three \emph{independent} amplitudes: one where the two neutrinos have opposite helicity, and two where the neutrinos have the same helicity, \emph{i.e}  $\mathcal{M}[L,R]$, $\mathcal{M}[R,R]$ and $\mathcal{M}[L,L]$. Both $\mathcal{M}[R,R]$ and $\mathcal{M}[L,L]$ must vanish when the momenta of the two neutrinos are identical due to the Pauli exclusion principle. Whereas for $\mathcal{M}[L,R]$ this is not required since the two neutrinos remain distinguishable by their helicity. The amplitudes $\mathcal{M}[L_1,R_2]$ and $\mathcal{M}[R_1,L_2]$ are not functionally independent (differing only by momentum interchange) but they represent distinct helicity states at a given $p_1$ and $p_2$ phase-space point. \\

The anti-symmetrization given in Option A satisfies each of these requirements.  In this set of amplitudes the two Dirac amplitudes that are anti-symmetrized are identical except for which is fermion and which is an anti-fermion. The two same-helicity amplitudes vanish when the two neutrino momenta are identical, and there is only one independent opposite-helicity amplitude as required.\\

Option B has a number of physics pathologies that make this anti-symmetrization  unacceptable to represent the Majorana amplitudes.
Firstly, there are two independent opposite-helicity amplitudes and not one. These two amplitudes are very different in magnitude: one has no helicity suppression, whereas the other is doubly helicity suppressed. There is only one same-helicity amplitude, and this amplitude does not vanish when the two neutrino momenta are identical -- violating the Pauli exclusion principle.  \\

Our conclusion here is that the anti-symmetrization given by Option A, on the left side of Fig. \ref{fig:AntiS}, is the only acceptable way to form Majorana neutrinos and that the $\mathcal{M}[L,R]$ amplitude is given by $L_1\overline{R}_2-\overline{L}_1R_2$, the $\mathcal{M}[R,R]$ amplitude by  $R_1\overline{R}_2-\overline{R}_1R_2$, and $\mathcal{M}[L,L]$ amplitude by  $L_1\overline{L}_2-\overline{L}_1L_2$. In the next section we explicitly calculate these Majorana amplitudes for $\Phi$ decay.

\subsection{Massive Majorana Neutrinos}
\label{sec:Maj}

To obtain the amplitudes for the case where the neutrinos are Majorana we have to anti-symmetrize the Dirac amplitudes, so that there can be no distinction between which neutrino is a fermion and which neutrino is an anti-fermion, leaving everything else unchanged. The correct way to do this follows from the left side of Fig.~\ref{fig:AntiS}, Option A. Below we list these amplitudes, after removing the same overall phase $(-e^{-i(\psi_1+\psi_2)})$ as in the Dirac case from each.\\

The {\it same helicity} Majorana amplitudes are given by
\begin{subequations}
\begin{align}
  {\cal M}[\, R_1, {R_2}\, ]  
 \equiv& 
   ~{\cal D}[\, R_1, \overline{R_2}\, ] -{\cal D}[\, R_2, \overline{R_1}\, ] =- {\cal M}[\, R_2, {R_1}\,  ]
\notag \\
 =& 
   + 2 \braketLL{\bar{k}}{\widehat{p}_2} \braketRR{\widecheck{p}_1}{k} ~\{ \braketLL{\widehat{p}_1}{\widecheck{p}_1} /m_\nu \}
\notag\\
 & 
   -2\braketLL{\bar{k}}{\widehat{p}_1}  \braketRR{\widecheck{p}_2}{k}  ~\{ \braketLL{\widehat{p}_2}{\widecheck{p}_2} /m_\nu \},
\\[2mm]
  {\cal M}[\, L_1, {L_2}\, ]  \equiv&~ {\cal D}[\, L_1, \overline{L_2}\, ] -{\cal D}[\, L_2, \overline{L_1}\, ]
 =
   - {\cal M}[\, L_2, {L_1}\,  ]\notag \\
 =&
   +2  \braketLL{\bar{k}}{\widecheck{p}_2} \braketRR{\widehat{p}_1}{k}~\{ \braketLL{\widecheck{p}_1}{\widehat{p}_1} /m_\nu \} \notag\\
 &
   -2 \braketLL{\bar{k}}{\widecheck{p}_1}  \braketRR{\widehat{p}_2}{k} 
   ~\{ \braketLL{\widecheck{p}_2}{\widehat{p}_2} /m_\nu \}.
 \end{align}
  \label{eq:MampS}
 \end{subequations}
 Note that these two amplitudes are odd functions of $1$ and $2$ ( \emph{i.e.} $\widehat{p}_1 \leftrightarrow 
 \widehat{p}_2$ and  $\widecheck{p}_1 \leftrightarrow \widecheck{p}_2$ ), and therefore vanish when $\widehat{p}_1 =\widehat{p}_2$ and $\widecheck{p}_1 =\widecheck{p}_2$. This follows from the Pauli exclusion principle: the two fermions cannot be in identical momentum and spin states.\\
 
 For the {\it opposite helicity}  Majorana amplitudes,
 \begin{subequations}
\begin{align}
{\cal M}[\, L_1, {R_2}\, ]  \equiv&~ {\cal D}[\, L_1, \overline{R_2}\, ] -{\cal D}[\, R_2, \overline{L_1}\, ] 
\notag \\
 =&
   +2 \braketLL{\bar{k}}{\widehat{p}_2} \braketRR{\widehat{p}_1}{k}  ~ \{ \braketLL{\widehat{p}_1}{\widecheck{p}_1} /m_\nu \}  
\notag  \\
&
   + \, 2 \braketLL{\bar{k}}{\widecheck{p}_1}  \braketRR{\widecheck{p}_2}{k}~\{ \braketLL{\widehat{p}_2}{\widecheck{p}_2} /m_\nu \},
   \\[2mm]
 {\cal M}[\, R_1, {L_2}\, ]  \equiv&~ {\cal D}[\, R_1, \overline{L_2}\, ] -{\cal D}[\, L_2, \overline{R_1}\, ] =- {\cal M}[\, L_2, {R_1}\,  ] 
 \notag \\
 =&
   -2 \braketLL{\bar{k}}{\widecheck{p}_2} \braketRR{\widecheck{p}_1}{k}  ~ \{ \braketLL{\widehat{p}_1}{\widecheck{p}_1} /m_\nu \}  
\notag  \\
&
   - \, 2 \braketLL{\bar{k}}{\widehat{p}_1}  \braketRR{\widehat{p}_2}{k}~\{ \braketLL{\widehat{p}_2}{\widecheck{p}_2} /m_\nu \},
 \end{align}
  \label{eq:MampO}
 \end{subequations}
Neither of these opposite helicity amplitudes are odd functions in $p_1$ and $p_2$, and so do not vanish when $p_1=p_2$. This is allowed as the two neutrinos have opposite helicity, so are not in identical states.
However, the two amplitudes in eq.~\eqref{eq:MampO} are related to one another as $ {\cal M}[\, R_1, {L_2}\, ]= -{\cal M}[\, L_1, {R_2}\, ](p_1 \leftrightarrow p_2)$, as expected since they are not independent amplitudes. There are only three independent amplitudes here: one for when both neutrinos are left-handed, one when both are right-handed and a single amplitude for when one is left handed and the other right handed, as required. In addition these are related to one another by appropriate $\widehat{p}_i \leftrightarrow \widecheck{p}_i$ interchanges, as this is equivalent to flipping the spin/helicity. \\

The amplitudes in eq.~\eqref{eq:Damps}, eq.~\eqref{eq:MampS}, and \eqref{eq:MampO} illustrate the power of the spinor formalism to study the neutrinos nature. Specially, it becomes trivial to see
that at the massless limit both Dirac and Majorana amplitudes are the same. No sum over spin neither integration over neutrino momentum variables are needed for such result to hold. From eq.~\eqref{eq:Damps} and in the helicity basis of eq.~\eqref{eq:helbasis}, the presence of $\widecheck p_i$ momentum, excluding the phases inside $\{\cdots\}$, in $\mathcal D[R_1, \overline R_2],\, \mathcal D[L_1, \overline L_2]$, and $\mathcal D[R_1, \overline L_2]$ shows only the Dirac amplitude $\mathcal D[L_1, \overline R_2]$ is non-zero for $m_\nu\rightarrow 0$. Combining this result with eq. \eqref{eq:MampS} and \eqref{eq:MampO},  the only amplitudes {\it not} containing a $\widecheck p_i$ momentum are the first term of $\mathcal M[L_1, R_2]$ and the second term of $\mathcal M[R_1, L_2]$. In the limit
 $m_\nu\rightarrow 0$, these two Majorana amplitudes go to the corresponding Dirac amplitude as follows: $\mathcal M[L_1, R_2] \rightarrow \mathcal D[L_1, \overline R_2]$ and $\mathcal M[R_1, L_2] \rightarrow  - \mathcal D[L_2, \overline R_1]$.  This result directly follows from the left-chiral nature of neutrino interactions in the SM.  \\

 The square of these Majorana amplitudes are given by 
 \begin{align}
|\, {\cal M}[\, R_1, {R_2}\, ]  \,|^2 & ~\varpropto ~ (k \cdot \widecheck{p}_1)(\bar{k}\cdot \widehat{p}_2)
+(k \cdot \widecheck{p}_2)(\bar{k}\cdot \widehat{p}_1) -\,{\rm Tr}[k \widecheck{p}_1 \widehat{p}_1
\bar{k} \widehat{p}_2 \widecheck{p}_2]/(4m_\nu^2) \notag \,, \\
|\, {\cal M}[\, L_1, {L_2}\, ] \,|^2 & ~\varpropto ~  (k \cdot \widehat{p}_1)(\bar{k}\cdot \widecheck{p}_2)
+(k \cdot \widehat{p}_2)(\bar{k}\cdot \widecheck{p}_1) -\,{\rm Tr}[k \widehat{p}_1 \widecheck{p}_1
\bar{k} \widecheck{p}_2 \widehat{p}_2]/(4m_\nu^2) \notag  \,,\\
|\, {\cal M}[\, L_1, {R_2}\, ]  \,|^2 & ~\varpropto ~  (k \cdot \widehat{p}_1)(\bar{k}\cdot \widehat{p}_2)
+ (k \cdot \widecheck{p}_2)(\bar{k}\cdot \widecheck{p}_1) -\,{\rm Tr}[k \widehat{p}_1 \widecheck{p}_1
\bar{k} \widehat{p}_2 \widecheck{p}_2]/(4m_\nu^2) \notag \,,\\
|\, {\cal M}[\, R_1, {L_2}\, ]  \,|^2 &~\varpropto ~ (k \cdot \widehat{p}_2)(\bar{k}\cdot \widehat{p}_1) +(k \cdot \widecheck{p}_1)(\bar{k}\cdot \widecheck{p}_2)
- \,{\rm Tr}[k \widecheck{p}_1 \widehat{p}_1
\bar{k} \widecheck{p}_2 \widehat{p}_2]/(4m_\nu^2) \,.
\label{eq:Msqed}
\end{align}
The ``slashes" on the momenta in the traces are omitted for readability.
 This set of four amplitudes squared has to be summed to obtain the full amplitude squared at each phase space point. 
All of these amplitudes squared are consistent with the left-chiral contribution to eq.(1) of Ma and Pantaleone~\cite{Ma:1989jpa}, where they calculate the process $e^+e^- \rightarrow NN $, where $N$ is a massive Majorana fermion. An explicit discussion of this may be found in Appendix~\ref{app:MP}.\\

The $1/m_\nu^2$ in the interference term, maybe a concern to the reader in the limit that $m_\nu \rightarrow 0$. However in the helicity basis for which this limit is smooth and using  $\widehat{p}$ and  $\widecheck{p}$ given in eq. \eqref{eq:helbasis}, it is clear that each of these interference terms are proportional to the neutrino mass squared, such as
\begin{align}
{\rm Tr}[k \widecheck{p}_1 \widehat{p}_1
\bar{k} \widecheck{p}_2 \widehat{p}_2]/m_\nu^2  \sim {\cal O}(\, |k|\,|\widecheck{p}_1|\, | \widehat{p}_1|\,
|\bar{k}|  \, | \widecheck{p}_2|\, |\widehat{p}_2|/m_\nu^2 \, = m_\nu^2\, |k|\, |\bar{k}| \,) \,,
\end{align}
where $|k|$ is the energy of the massless momentum $k$. 
As there is no Lorentz $\gamma \, (= E/m )$ factor's for the neutrinos in the magnitude of this trace, these trace factors go to zero as $m_\nu \rightarrow 0$. 
This is exactly the DMCT for each helicity configuration. \\

Summing over all the spin combinations for the square of the Majorana amplitudes, we have
\begin{align}
\sum_{spins} |\, {\cal M}[\, 1,2\, ]  \,|^2 & \propto  (k \cdot {p}_1)(\bar{k}\cdot {p}_2)
+(k \cdot {p}_2)(\bar{k}\cdot {p}_1) - m_\nu^2 (k \cdot \bar{k})  \,.
\label{eq:totM}
\end{align}
This is the spin summed full amplitude squared for the decay of the $\Phi$ assuming that the neutrinos are Majorana particles and that lepton number is  not conserved. Note that if we integrate this amplitude-squared over the full phase space for $p_1$ and $p_2$, we further divide by two so as to not double count. Except for the last term, $ m_\nu^2 (k \cdot \bar{k})$, which goes to zero when $m_\nu \rightarrow 0$, this is identical to the expression obtained for the Dirac neutrinos\footnote{The omitted constant factors are identical for the Dirac and Majorana cases, so comparing eq.~\eqref{eq:totD} and~\eqref{eq:totM} is exact.}, eq. \eqref{eq:totD}. This confirms the DMCT for this process and is in agreement with the general arguments given by Akhmedov and Trautner in ref.~\cite{Akhmedov:2024qpr}. \\

The only phase space point where this summed  Majorana amplitude squared is zero is when the neutrinos are at rest in the  $\Phi$  rest frame as the two massless charged leptons must have opposite momenta, 
$ (k \cdot \bar{k}) = 2 |k | |\bar{k} | $, and the spin of the two Majorana neutrinos must be in the direction of the negative charged lepton by angular momentum conservation and therefore the two neutrinos are in the same momentum and spin state.  When $p_1=p_2$ and the two neutrinos are relativistic in the $\Phi$ rest frame, the fractional change to squared amplitude from the interference term is of ${\cal O}(1/\gamma^2)$, where $\gamma m_\nu$ is the energy of the neutrinos in this frame.
We will extensively compare the spin summed Dirac and Majorana amplitudes squared in the next section.

\subsection{Comparison of the Dirac and Majorana cases }
\label{sec:Pheno}
\begin{figure}[t!]
      \centering
      \includegraphics[width=0.55\linewidth]
      {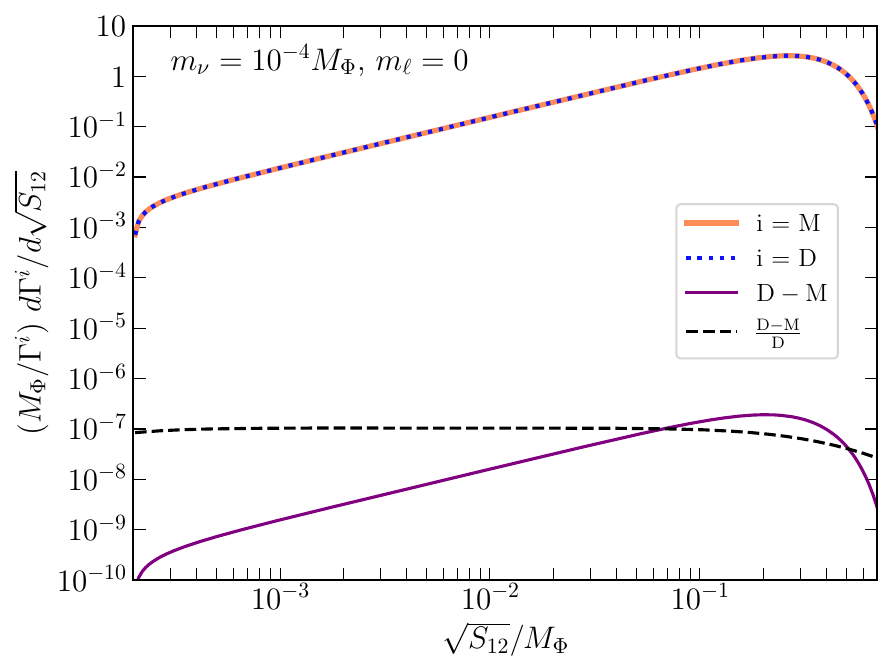}
      \caption{The differential decay rate for $\Phi$ versus the square root of the invariant mass of the two neutrinos, $\sqrt{S_{12}}$, for a neutrino mass of $m_\nu = 10^{-4} M_\Phi$ and massless charged leptons.  The Dirac distribution is the red line and the Majorana distribution is the cyan-dashed line. The purple line is the difference between Dirac and Majorana distribution    and the black dotted line this difference divided by the Dirac distribution. Note, the roughly flat relative  difference between the Dirac and Majorana distributions for all values of $\sqrt{S_{12}}$. The choice  $m_\nu = 10^{-4} M_\Phi$ is just representative and shows the relevant physics.   }
      \label{fig:dGammads12_massless}
  \end{figure}
  
Angular correlations for the decay of $\Phi$ are determined by the matrix element squared, as given for the Dirac and Majorana cases in eq. \eqref{eq:totD} and \eqref{eq:totM}. To summarize, these are given by
\begin{align}
(k \cdot {p}_1)(\bar{k}\cdot {p}_2)
+(k \cdot {p}_2)(\bar{k}\cdot {p}_1)  - 
\left\{\begin{array}{ll}
0 & \quad\text{[Dirac]}\\[3mm]
m_\nu^2 (k \cdot \bar{k}) \quad\quad & \text{[Majorana]}
\end{array}
\right.
\,,
\label{eq:totB}
\end{align}
where the same common factor has been suppressed for both cases.  Note that here we consider massless charged leptons, but address the impact of their masses in Section~\ref{sec:CLMass}. \\

Following from eq.~\eqref{eq:totB}, the difference between the Dirac and Majorana amplitudes squared, divided by the Dirac amplitude squared, is given by
\begin{align}
\frac{\text{D}-\text{M}}{\text{D}}  
  &\equiv
    \frac{ m_\nu^2 (k \cdot \bar{k}) }{ (k \cdot p_1)(\bar{k} \cdot {p}_2) + (k \cdot p_2)(\bar{k} \cdot {p}_1)}
 \notag     \\[2mm] &
    =\biggr[ \frac{1}{ \gamma_{p_1}  \gamma_{p_2}} \biggr]~\biggr( \frac{(1-c_{k\bar{k}} )} { (1-\beta_{p_1} c_{p_1k})(1-\beta_{p_2} c_{p_2\bar{k}}) 
    +(1-\beta_{p_1} c_{p_2k})(1-\beta_{p_2}c_{p_1\bar{k}})} \biggr)
\label{eq:DifferenceDM_massless}
  \end{align}
  where $c_{q q'} =\cos \theta_{q q'}$ is the cosine of the angle between momenta $q$ and $q'$, and $\beta_q, \gamma_{q}$ are the relativistic momentum and Lorentz factors for the massive neutrino with momenta $q$. \\
  
  In the rest frame of the $\Phi$, the factor $(\cdots)$ is of order one. The factor in $[\cdots]$ is \emph{only} of order one when both neutrinos are non-relativistic, \emph{i.e.} both $\gamma_{p_1}, \gamma_{p_2}$  of order one. One may be tempted to consider this in a Lorentz-invariant way by requiring
  the invariant mass of the two neutrinos, $S_{12}$, to be close to threshold:
    \begin{align}
         S_{12} \equiv (p_1+p_2)^2 \approx  4m_\nu^2. \label{eq.threshold} 
        \end{align}
 Accordingly, the difference between Dirac and Majorana neutrinos is only significant in the region of phase space where one can boost to a frame where both neutrinos are non-relativistic. In Fig. \ref{fig:dGammads12_massless}, we show both the Dirac and Majorana differential decay rates, and their relative difference, as a function of the square-root of $S_{12}$. Note that the difference between the two is \emph{always} suppressed by $\sim {\cal O}(m_\nu^2/M^2_\Phi)$. However, eq.~\eqref{eq.threshold} is only a necessary but not a sufficient condition. To see this, consider eq.~\eqref{eq:DifferenceDM_massless} when $p_1=p_2$. This satisfies the $S_{12}$ requirement, but does not guarantee that there is a significant difference between Dirac and Majorana neutrinos because, at this point, $\gamma_1 \gamma_2$ can still be $\gg 1$.  \\

\begin{figure}[t!]
      \centering
         \includegraphics[scale=0.5]{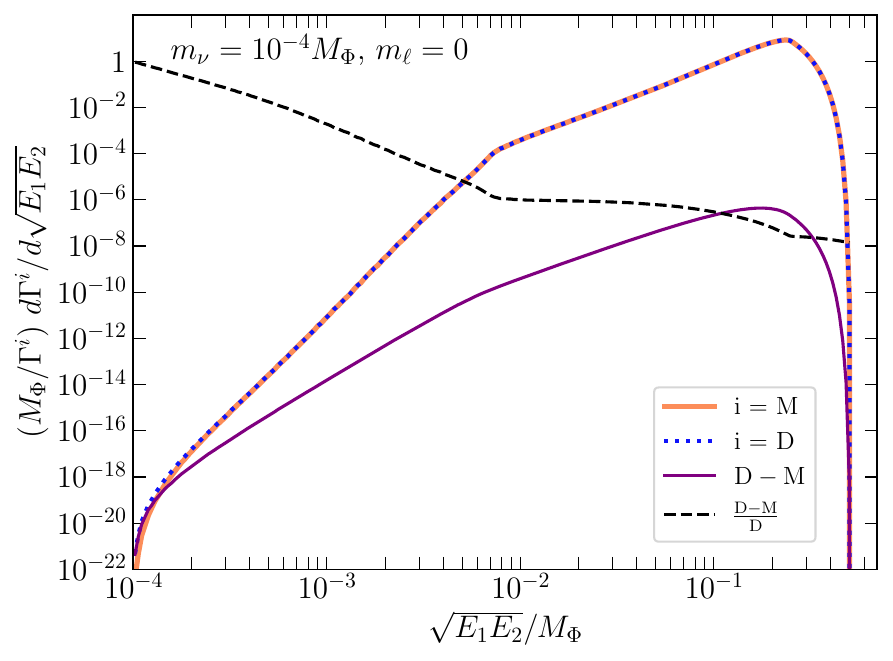}
             \hspace{5mm}
             \includegraphics[scale=0.5]{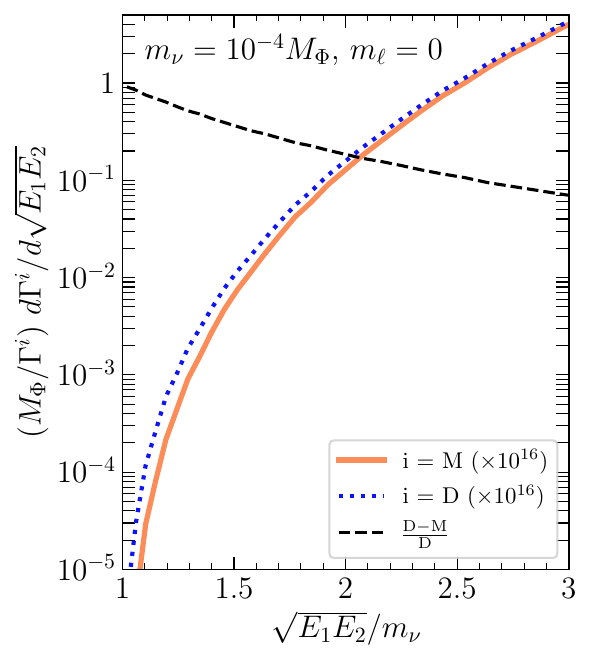}
      \caption{\textbf{Left panel:} the differential decay rates of $\Phi$ versus the square root of the product of the energy of the two neutrinos, $\sqrt{E_1E_2}$, in the rest frame of $\Phi$, for a neutrino mass of $m_\nu = 10^{-4} M_\Phi$ and massless charged leptons.
      The Dirac distribution is the red line and the Majorana distribution is the cyan-dashed line. The purple line is the difference between Dirac and Majorana distribution and the black dotted line this difference divided by the Dirac distribution. Note, the relative  difference between the Dirac and Majorana distributions only becomes significant for small  values of $\sqrt{E_1E_2}/M_\Phi$ as this is when both neutrinos are non-relativistic in the $\Phi$ rest frame. This region of phase space is suppressed by powers of $(m_\nu/M_\Phi)$.  \textbf{Right panel:} we zoom in on the $\sqrt{E_1E_2} \sim m_\nu$ region. When the neutrinos are produced at rest, the Majorana process is zero by angular momentum conservation and the Pauli exclusion principle, whereas the Dirac process is non-zero. Therefore (D$-$M)/D~$=1$ when $\sqrt{E_1E_2}= m_\nu$.}
      \label{fig:dGammadE1E2_massless}
  \end{figure}

The correct Lorentz-invariant condition is that 
\begin{align}
(Q \cdot p_1)(Q  \cdot p_2) \sim {\cal O}(m_\nu^2M^2_\Phi) \,,
\label{eq:NonR}
\end{align} 
recalling that $Q$ is the four-momentum of the decaying $\Phi$. That is, in the rest frame of $\Phi$, both neutrinos must be non-relativistic. 
The fraction of phase space that satisfies these conditions is tiny and of ${\cal O}(4m_\nu^2/M^2_\Phi)$.\footnote{For a scattering process rather than the decay of a particle, the same condition will apply with $Q$ replaced by the sum of the momenta of the incoming particles, \emph{i.e.} in the zero three-momentum frame the two neutrinos must be non-relativistic.}  In Fig.~\ref{fig:dGammadE1E2_massless} we show both the Dirac and Majorana differential decay rates, and their relative difference, as a function of the square-root of $\sqrt{E_1 E_2} \equiv \sqrt{(Q \cdot p_1)(Q  \cdot p_2)}/M_\Phi$ in the rest-frame of the scalar $\Phi$. As expected, the \emph{relative} difference grows with decreasing $\sqrt{E_1 E_2}$ and reaches a maximum at threshold for both neutrino's production at rest, when $\sqrt{E_1 E_2} = m_\nu$, where the Majorana decay rate is zero.

  \subsection{Adding Masses to the Charged Leptons}
\label{sec:CLMass}

Here we consider the impact of assigning the charged-leptons nonzero masses on our result in Section~\ref{sec:Pheno}. The Dirac amplitudes for massive charged leptons may be obtained by making the following replacements in eq.~\eqref{eq:Damps},
\begin{align}
 \ketR{k} ~ & \mapsto ~ \left\{ 
 \begin{array}{lr}
+  \ketR{\widecheck{k}}  
& \text{R lepton} \\[2mm]
+ \ketR{\widehat{k}} 
& \text{L lepton} 
 \end{array}
   \right. \,, \quad \quad \quad    \braL{\overline{k}} ~  \mapsto ~  \left\{ 
 \begin{array}{lr}
+  \braL{~\widehat{\overline{k}}}  & \text{R anti-lepton} \\[2mm]
- \braL{~\widecheck{\overline{k}}}  & \text{L anti-lepton} 
 \end{array}
   \right.\, ,
   \label{eq:massleprepl}
\end{align}
which follows from using the spinors from eq.~\eqref{eq:spinors} and eq.~\eqref{eq:DiracLine2}.
The common phase introduced by the massive charged fermions features as an overall factor for both Dirac and Majorana neutrinos, and therefore can be consistently factored out and has no physical impact. The all spin summed Dirac amplitude squared is then given by
 \begin{align}
   \sum_\text{all~spins}  |\,{\cal D}[\,1 \, ,  \overline{2}\, ]\, |^2  +  |\,{\cal D}[\,2 \, , \overline{1}\, ]\, |^2 & \propto 
   (k \cdot p_1)(\bar{k} \cdot {p}_2) + (k \cdot p_2)(\bar{k} \cdot {p}_1) \,,
   \label{eq:totD2}
 \end{align}
 now with $k^2=\overline{k}^2 =m^2_l$ instead of zero.\\

Similarly for Majorana case, one has to make the replacements from eq.~\eqref{eq:massleprepl} in eq. \eqref{eq:MampS} and eq.~\eqref{eq:MampO}. The square of the helicity amplitude with $(L\overline{R})_{l^-l^+}$ for the charged leptons and $(LR)_{\nu\nu}$ for the neutrinos is given by
\begin{align}
|\,{\cal M}[\,L_k \overline{R}_{\bar{k}}\, | \, L_1 R_2\, ] \, |^2 & \propto 
   \left( \widehat{k} \cdot \widehat{p}_1 \right) \left(\widehat{\overline{k}} \cdot \widehat{p}_2 \right) + \left(\widehat{k} \cdot \widecheck{p}_2 \right) \left(\widehat{\overline{k} }\cdot \widecheck{p}_1 \right)
   -\,{\rm Tr}\left[\, \widehat{k}\, \widehat{p}_1 \widecheck{p}_1\,
\widehat{\overline{k}} \,\widehat{p}_2 \widecheck{p}_2 \, \right]/ (4m_\nu^2) \,,
   \end{align}
and similarly for the other helicity amplitudes. If one sums over the charge lepton configurations, this recovers the expression in eq.~\eqref{eq:Msqed}, since each term is linear in both $k$ and $\overline{k}$. \\

The all-spin summed Majorana amplitude squared is again given by
\begin{align}
\sum_\text{all spins} |\, {\cal M} \,|^2 & \propto  (k \cdot {p}_1)(\bar{k}\cdot {p}_2)
+ (k \cdot {p}_2)(\bar{k}\cdot {p}_1) -  m_\nu^2 (k \cdot \bar{k})  \,.
\label{eq:totM2}
\end{align}
again with $k^2=\overline{k}^2 =m^2_l$ rather than zero.  So for both the Dirac and Majorana amplitudes squared, the charged lepton mass does not appear \textit{explicitly} but only \textit{implicitly} in the relativistic momentum of the two charged leptons. \\

 \begin{figure}[t!]
      \centering
       \includegraphics[scale=0.50]
       {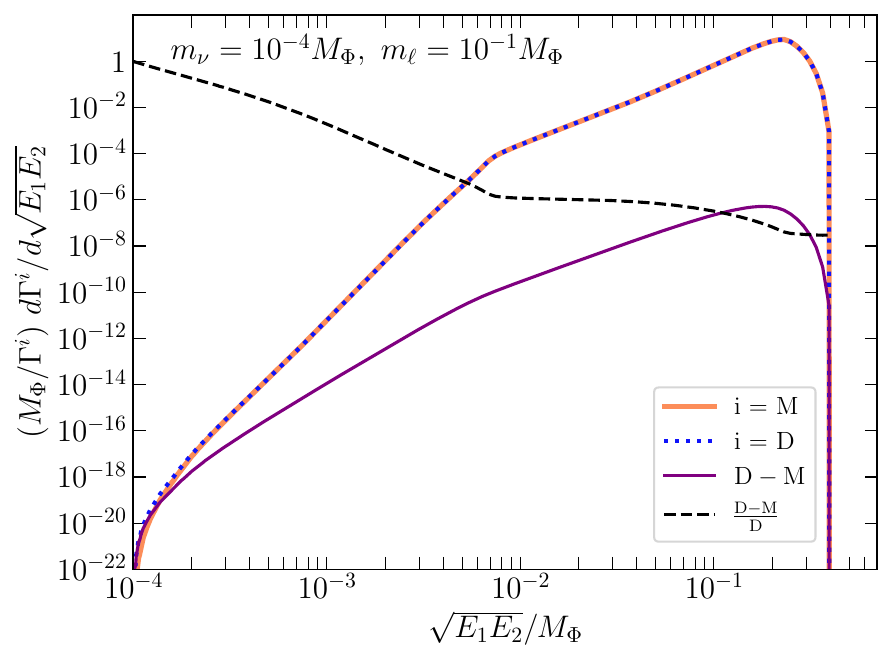}
      \caption{ The differential decay rates of $\Phi$ versus the square root of the product of the energy of the two neutrinos, $\sqrt{E_1E_2}$, in the rest frame of $\Phi$, for a neutrino mass of $m_\nu = 10^{-4} M_\Phi$ and charged lepton mass $m_l = 10^{-1} M_\Phi$. 
      The Dirac distribution is the red line and the Majorana distribution is the cyan-dashed line. The purple line is the difference between Dirac and Majorana distribution and the black dotted line this difference divided by the Dirac distribution. Note, the relative  difference between the Dirac and Majorana distributions only becomes significant for small  values of $\sqrt{E_1E_2}/M_\Phi$ as this is when both neutrinos are non-relativistic in the $\Phi$ rest frame. The most significant difference between the charged lepton massless, Fig. \ref{fig:dGammadE1E2_massless}, and massive is the position of the kinematic end point at large $\sqrt{E_1E_2}/M_\Phi$ and when $\sqrt{E_1E_2}=m_\nu$ the ratio (D$-$M)/D$\approx 1-2(m_\ell/M_\Phi)^2= 0.98$, as expected from eq. \eqref{eq:DmMoverDmassive}.
      }
      \label{fig:dGammadE1E2_massive}
  \end{figure}

 Therefore, the difference between Dirac and Majorana amplitudes in eq.~\eqref{eq:DifferenceDM_massless} becomes
  \begin{align}
\frac{\text{D}-\text{M}}{\text{D}}    &=\biggr[ \frac{1}{ \gamma_{p_1}  \gamma_{p_2}} \biggr]~\biggr( \frac{(1- \beta_k \beta_{\bar{k}} c_{k\bar{k}} )} { (1- \beta_{p_1} \beta_k c_{p_1 k})(1-\beta_{p_2}  \beta_{\bar{k}} c_{p_2\bar{k}}) 
    +(1-\beta_{p_1} \beta_k c_{p_2 k})(1-\beta_{p_2}  \beta_{\bar{k}} c_{p_1\bar{k}})} \biggr)
\label{eq:DmMoverDmassive}
  \end{align}
The only difference between this and eq. \eqref{eq:DifferenceDM_massless}, is that $\beta_k $ and $\beta_{\bar{k}} $ have been introduced. As both  $\beta_k\approx 1 $ and $\beta_{\bar{k}} \approx 1$ in most regions of phase space in the rest frame of the $\Phi$, the effect of adding masses to the charged leptons is minor and does not change the physics conclusion that the DMCT is valid.\\

Note that unlike when the charged leptons were treated as massless, the Majorana amplitude squared does \emph{not} exactly vanish  when the two neutrinos are at rest in the $\Phi$ rest frame. This is because there is a small contribution from when the helicity configuration of the two charged leptons is $L\overline{L}$ and  $R\overline{R}$. In both of these configurations, the two neutrinos must have opposite spin and are therefore not in identical states. Thus the amplitude squared is not forced to be zero by the Pauli exclusion principle, as was the case for massless charged leptons. Nevertheless, the magnitude of the amplitude squared is tiny here due to the additional helicity suppression for the $L\overline{L}$ and  $R\overline{R}$ charged lepton configuration. At threshold, (D$-$M)/D $= 1-2m^2_\ell/(M_\Phi-2m_\nu)^2$ which equals 1 for massless charged leptons, and goes to $1/2$ as the mass of the charged leptons approaches its kinematic upper limit, $m_\ell = M_\Phi/2 -m_\nu$.\\

In Fig.~\ref{fig:dGammadE1E2_massive} we show the decay distribution obtained by integrating the spin-sum matrix elements squared for Dirac and Majorana neutrinos considering a non-zero charged lepton mass. The decay distribution shape are roughly the same as the ones for the massless charged fermion. The inclusion of small charged fermion masses mostly changes the near-threshold curves and makes the relative difference between Dirac and Majorana curves to be slightly smaller than in the $m_\ell = 0$ case since for massive charge leptons the Majorana decay rate is non-zero at the threshold point.

\section{ \texorpdfstring{Application to $B_0$ decay}{Application to B0 decays}}
\label{sec:b0decay}

Here we turn our attention to the application of our results to leptonic neutral-meson decays. In ref.  \cite{Kim:2021dyj}, the authors discuss the possibility of using particle correlations in the decays of $B_0$ to two charged leptons and two neutrinos,
  \begin{align}
  B_0 \rightarrow  \mu^-  \, \nu_1 ~ \mu^+ \, \nu_2  \,,
  \label{eq:B0decay}
  \end{align}
  to distinguish between Majorana and Dirac neutrinos.
  The weak interaction part for this decay is identical to the decay of $\Phi$ discussed in Sec.  \ref{sec:Scalar}. The $B_0$ decay has additional complications due to strong interactions, however these do not effect the correlations of the leptons. Therefore, the $\Phi$ decay discussed in Sec.  \ref{sec:Scalar} is a good, simple, model to understand the leptonic structure of $B_0$ decay, in particular to understand the anti-symmetrization of the Dirac amplitudes to obtain the Majorana amplitudes.\\

Unfortunately, in ref.~\cite{Kim:2021dyj} the authors of are not very explicit about how they construct the Dirac amplitudes, and hence the Majorana amplitudes, especially as it applies to the spin/helicity configurations of the two neutrino states. The most direct reference is in Fig. 1 of ref.~\cite{Kim:2021dyj} , where 
\begin{align}
{\cal M}^D(p_1, p_2) \sim {\cal D}[~L_2, \overline{R}_1]  
\,, 
\label{eq:KD1}
\end{align}
and the Dirac amplitude squared is given by
\begin{align}
| {\cal M}^D(p_1, p_2)|^2 = | {\cal D}[~L_2, \overline{R}_1]|^2 \sim (k \cdot p_2)(\bar{k} \cdot p_1)
\,. 
\end{align}
Note that this is assumes that $p_1$ is the anti-neutrino momentum. The authors have not included the possibility that $p_2$ is the anti-neutrino, which would add the term
\begin{align}
 |{\cal D}[~L_1, \overline{R}_2] |^2
 \sim (k \cdot p_1)(\bar{k} \cdot p_2)\, 
\end{align}
to the Dirac matrix-element squared.
To make a fair comparison with the Majorana case, as the neutrinos are {\it not} observed, it is imperative that \emph{both} momentum assignments be the anti-neutrino are summed to construct the Dirac matrix-element squared as
\begin{align}
| {\cal M}^D(p_1, p_2)|^2_\text{FULL} \sim 
 (k \cdot p_1)(\bar{k} \cdot p_2) +(k \cdot p_2)(\bar{k} \cdot p_1) \,\label{eq:FullForKim}
\end{align}
consistent with our eq.~\eqref{eq:totD}.\\

The construction of the Majorana amplitude given in ref.~\cite{Kim:2021dyj} is
\begin{align}
{\cal M}^M(p_1,p_2) & \sim {\cal M}^D(p_1,p_2) - {\cal M}^D(p_2,p_1) \,, \notag \\
 & \sim ~ {\cal D}[~L_2, \overline{R}_1 \,] - {\cal D}[~L_1, \overline{R}_2 \,]
\end{align}
Note that if $p_1=p_2$, this amplitude vanishes.  Furthermore, if one squares this Majorana amplitude they arrive at
\begin{align}
|{\cal M}^M(p_1,p_2)|^2 &  \sim |{\cal M}^D(p_1,p_2) - {\cal M}^D(p_2,p_1)|^2 \\
 & \sim (\bar{k} \cdot p_1)
(k \cdot p_2) + (\bar{k} \cdot p_2)(k \cdot p_1) +\text{ (interference terms)} \,,
 \end{align}
 These interference terms must be sizeable in order to make the amplitude squared vanish when $p_1=p_2$, and therefore they cannot be proportional to powers of the neutrino mass. Except for the interference terms, this matrix element squared is the same as the full Dirac matrix element squared in eq.~\eqref{eq:FullForKim}, and thus the interference terms violate the DMCT. However, this construction of the Majorana amplitude reflects the $\overline{R}_1 L_2 - L_1 \overline{R}_2 $ of Option B discussed in Sec. \ref{sec:AntiS}, which has numerous physics pathologies. \\

In ref.~\cite{Kim:2023iwz}, a more explicit definition of the Dirac amplitude is given (eq. (28) of ref.~\cite{Kim:2023iwz})  as the coherent helicity-sum of the Dirac helicity amplitudes,
 \begin{align}
{\cal M}^D(p_1, p_2) \sim \sum_{\lambda_1 \lambda_2} {\cal D}[ ~  \lambda_2, \overline{\lambda}_1)\,] 
\,,
\label{eq:KD2}
\end{align}
where $\lambda_i$ is the helicity of the $i$-th neutrino.
This is a very unusual prescription for forming the Dirac amplitude as the helicity sum is always performed after squaring the individual amplitudes, see for example ref.~\cite{Halzen-pg-120}.\\

Conceptually, this definition in eq.~\eqref{eq:KD2} from ref.~\cite{Kim:2023iwz} is very different from eq. \eqref{eq:KD1}, the definition given in ref.~\cite{Kim:2021dyj}. However, from a numerical 
point of view the difference is small, because one term in the helicity sum dominates while the other three are helicity suppressed -- following from the left-chiral nature of the weak interactions, and the smallness of the neutrino mass. \\

Following from ref.~\cite{Kim:2023iwz}, the Dirac matrix-element squared is given by
\begin{align}
| {\cal M}^D(p_1, p_2) |^2  \sim \left| \sum_{\lambda_1 \lambda_2} {\cal D}[\, \lambda_2, \overline{\lambda}_1)\,] \right|^2 \,.
\label{eq:KD2sq}
\end{align}

Note, in particular, that in eq.~\eqref{eq:KD2sq} the helicity sum is performed before squaring. The canonical way to do this sum is as
\begin{align}
 \sum_{\lambda_1 \lambda_2}  \left|{\cal D}[\,\lambda_2, \overline{\lambda}_1)\,] \right|^2 \,.
\label{eq:KD2sqC}
\end{align}
The latter guarantees that the matrix-element squared yields the same result independent of which spin axis is used for the neutrinos. This follows from the completeness relations given in Appendix~\ref{sec:Spinors}. Therefore, using the Dirac amplitude in eq.~\eqref{eq:KD2} would give different results depending on the neutrinos spin axis, and thus is not physical.\footnote{A further argument against this coherent, rather than incoherent, sum is given in the Appendix \ref{sec:phasechange}.} 
Also note again that this does not include the case where $p_2$ is the momentum of the anti-neutrino. \\

The Majorana amplitude in ref.~\cite{Kim:2023iwz} is constructed as
\begin{align}
{\cal M}^M(p_1,p_2) & \sim {\cal M}^D(p_1,p_2) - {\cal M}^D(p_2,p_1) \,, \notag \\
 & \sim  \sum_{\lambda_1 \lambda_2}  {\cal D}[\, \lambda_2, \overline{\lambda}_1\,] 
 -  \sum_{\lambda_1 \lambda_2} {\cal D}[\, \lambda_1, \overline{\lambda}_2\,].
\label{eq:KM2}
\end{align}
Note that the Majorana matrix-element in eq.~\eqref{eq:KM2} exactly vanishes when $p_1=p_2$: for each term in the first sum there is an identical term in the second. The matrix-element squared is given by
\begin{align}
|{\cal M}^M(p_1,p_2) |^2 
 & \sim  \left|  \sum_{\lambda_1 \lambda_2} \biggr( {\cal D}[\,\lambda_2, \overline{\lambda}_1)\,] 
 -  {\cal D}[ ~  \lambda_1, \overline{\lambda}_2)\,]
 \biggr) \right|^2.
\label{eq:KM2sq}
\end{align}
This suffers from the same malady as the Dirac amplitude, in that the answer will depend on the spin projections, whereas a physical answer should be independent of this choice. Numerically, note that this amplitude squared will again be dominated by the $\overline{R}_1 L_2 - L_1 \overline{R}_2$ from Option B disputed in Sec. \ref{sec:AntiS}.\\

We conclude that, in ref.~\cite{Kim:2021dyj} and \cite{Kim:2023iwz}, either they have incorrectly used the coherent sum of the spin Dirac amplitudes, or they have used the unphysical Option B for the anti-symmetrizing the Dirac amplitudes to construct the Majorana amplitudes. 
{Following instead from the results of Section~\ref{sec:Scalar}, the difference between Dirac and Majorana neutrinos is only significant when the two neutrinos are nearly at rest in the $B_0$ rest frame. This represents a fraction of phase space suppressed by powers of $(m_\nu/M_{B_0})$, thereby satisfying the DMCT and in agreement with the general arguments given by Akhmedov and Trautner in ref.~\cite{Akhmedov:2024qpr}.

\section{Beyond Standard Model Interactions}
\label{sec:BSM}

The spinor-helicity techniques introduced for application to Dirac and Majorana neutrinos may be extended also to Beyond Standard Model (BSM) scenarios. We describe an example of this here, in models with an additional right-handed $W$-boson or $Z^\prime$-boson. As was emphasized in ref.~\cite{Akhmedov:2024qpr}, the DMCT is only valid when the interactions of the neutrinos are left-chiral, as in the SM. For example, ref.~\cite{Rodejohann:2017vup} the authors consider observables for discriminating between Dirac and Majorana neutrinos with general interactions.\\

If we add a right-handed $W$-boson, then the Dirac amplitude $(\overline{L}_1 R_2)_{\nu\nu}$ is no longer helicity-suppressed by
the mass of the neutrinos, but is suppressed by the ratio of the $W$-boson masses squared. Therefore this amplitude, when added with  $(L_1 \overline{R}_2)_{\nu \nu}$ from the left handed W-boson to form the Majorana amplitude,  could give a sizable interference when squared.  However, at least for the $\Phi$ decay discussed in Section~\ref{sec:Scalar}, there is a further suppression of the interference from the charged leptons. For the right-handed $W$-boson, the  charged leptons are dominantly produced in $(R\overline{L})_{l^-l^+}$,  whereas for left-handed $W$-bosons the charged leptons are in $(L\overline{R})_{l^-l^+}$ state. Therefore, if the charged leptons are massless there is no interference between these Dirac states. For charged leptons with non-zero mass, this will lead to a suppression which is charged lepton mass dependent.  For $e^+e^-$ this suppression will be significant, whereas for $\tau^+ \tau^-$ it is not so significant. The techniques introduced here could easily be used to calculate the full matrix amplitude square including a right-handed $W$-boson if needed, however the discussion here highlights the important features.\\

For a process for which the neutrino pair are produced by a SM $Z$-boson in the  $(L_1 \overline{R}_2)_{\nu \nu}$, and $Z^\prime$ in the $(\overline{L}_1 R_2)_{\nu\nu}$, without a corresponding charged lepton pair, it may be possible to avoid this charged lepton mass suppression of the inference term. However there will still be a suppression associated with the couplings and mass of the $Z^\prime$-boson  compared with those of the $Z$-boson.

\section{Conclusions}
\label{sec:Concl}

In this work, we introduce the use of massive-spinor helicity methods to processes with massive neutrinos, in the Standard Model interaction framework. We explore the decay of a light scalar to two oppositely charged leptons and two neutrinos to assess whether quantum statistics can be used in this system to distinguish the nature of neutrinos as Dirac or Majorana. A measurable distinction between Dirac and Majorana neutrinos in this process would constitute a violation of the Dirac-Majorana confusion theorem~(DMCT). \\

Using these techniques, we calculated the helicity amplitudes for when the neutrinos are a Dirac particle and antiparticle pair, and also when they are a pair of Majorana neutrinos and thus there can be no particle-antiparticle distinction. The anti-symmetrization of the Dirac helicity amplitudes to form the Majorana amplitudes is explored in detail. A key result is the identification of the correct physical method for this anti-symmetrization, which when applied to the two-neutrino final state involves the following: 
\begin{enumerate}
    \item For a Dirac amplitude which is \underline{not helicity suppressed}, such as $L_1\overline{R}_2$, the anti-symmetrized is with a Dirac amplitude that is \underline{doubly  helicity suppressed}, $\overline{L_1}R_2$, and similarly vice versa;
    \item For a Dirac amplitude that is \underline{singly helicity suppressed}, such as $L_1\overline{L}_2$, the anti-symmetrized is with another \underline{singly suppressed} Dirac amplitude,  $\overline{L_1}L_2$.
    \end{enumerate}
This anti-symmetrisation corresponds to the simple interchange between fermion and antifermion for the neutrinos, without any additional spin flip. The helicity suppression is generated by the left-chiral nature of neutrino interactions in the Standard Model. Both above configurations imply that when the Majorana amplitudes are squared, the interference terms are also helicity suppressed and go to zero as the neutrino mass goes to zero: consistent with the DMCT. This argument may be straightforwardly generalized to final-states with more than two neutrinos. Our results are in agreement with the general arguments presented in ref.~\cite{Akhmedov:2024qpr}, and counter to the claims in refs.~\cite{Kim:2021dyj, Kim:2023iwz, Kim:2024tsm}.\\

Consistency with the DMCT is tied to the left-chiral interactions of neutrinos in the Standard Model framework. The techniques introduced in this work may be applied to BSM scenarios  for which quantum statistics may yield observable distinction between Dirac and Majorana neutrinos. Furthermore, our results highlight the relative simplicity and transparency to underlying physics in calculations with spinor-helicity techniques when  contrasted with traditional trace techniques. \\

\newpage

\appendix

\section{Technical Appendices}

\subsection{A Full Set of Spinors}
\label{app:morespinors}

  \begin{table}[t]
 \begin{center}
 \hspace*{-1cm} \begin{tabular}{|c|c|c|}
 \hline&& \\[-1mm]
  	{\bf SPINORS:} & Outgoing  & Incoming \\[2mm]
	\hline 
	&& \\
	Fermion & $ \begin{array}{rl} 
	   \overline{U}_\uparrow (p, s)  &=~~e^{-i\psi} \,\braL{\widehat{p}} + e^{i\psi} \, \braR{\widecheck{p}}   \\[2mm]
	\overline{U}_\downarrow (p,  s)   &= - e^{- i\psi }\, \braL{\widecheck{p}} + e^{i\psi}  \,\braR{ \widehat{p}}  \\[2mm]
	\end{array} $
	& 
	$\begin{array}{rl}
	 U_\uparrow (p, s) &= ~~e^{i\psi} \, \ketR{\widehat{p}} + e^{-i\psi}\,  \ketL{\widecheck{p}}   \\[2mm]
	 U_\downarrow (p, s) &= - e^{i\psi} \, \ketR{\widecheck{p}} +e^{-i\psi}  \,\ketL{\widehat{p}}  \\[2mm]
	\end{array} $\\
	\hline
	&& \\
	$\begin{array}{c}
	\text{Anti-}\\\text{Fermion}  \end{array} $
	 & $ \begin{array}{rl}
	 V_\uparrow (p, s)  &= ~~ e^{i\psi} \, \ketR{\widecheck{p}} + e^{-i\psi}\, \ketL{\widehat{p}}  \\[2mm]	
	 V_\downarrow (p, s)  &= ~~ e^{i\psi} \, \ketR{\widehat{p}} -e^{-i\psi} \, \ketL{\widecheck{p}}   \\[2mm]
	\end{array}$
	&
	$\begin{array}{rl}
	\overline V_\uparrow(p, s)   &= ~~ e^{-i\psi} \, \braL{\widecheck{p}} + e^{i\psi} \,\braR{\widehat{p}}  \\[2mm]
	\overline V_\downarrow(p,  s) &= ~~ e^{-i\psi}\, \braL{ \widehat{p}}  - e^{i\psi }\,  \braR{\widecheck{p}} \\[2mm]
	\end{array}$ \\
	\hline
	\end{tabular}
	\label{tab:Dspinors}
	\caption{Dirac spinors for a massive fermion with four-momentum $p$, mass $m$ ($p^2=m^2$), and spin four-vector $s$, 
	( $s^2=-1$ and  $p\cdot s =0$), we define two massless four-momenta as follows:
$\widehat{p} \equiv (p+ms)/2 \quad \& \quad \widecheck{p} \equiv (p-ms)/2 \,, $
then $\widehat{p}\,^2=0=\widecheck{p}\,^2$.  The phase, $\psi(p,s)$ is momentum and spin axis dependent is given by  
$\psi \equiv \frac1{2} \text{Arg}(  \braketLL{\widehat{p}}{\widecheck{p}}/m ) $. More on this phase is given in Table \ref{tab:phases}.
 }
	\end{center}
	\end{table}

       \begin{table}[b]
 \begin{center}
  \begin{tabular}{|cc|c|}
  \hline
  Phase Factor & & (Phase Factor)$^*$ \\[2mm]
   \hline && \\[-2mm]
  $ \braketLL{\widehat{p}}{\widecheck{p}}/m  \equiv e^{+i2\psi}$ &\quad &
  $ \braketRR{\widecheck{p}}{\widehat{p}}/m  = e^{-i2\psi} $  \\
  $ \braketLL{\widecheck{p}}{\widehat{p}}/m   = - e^{+i2\psi}$ &&
  $ \braketRR{\widehat{p}}{\widecheck{p}}/m  = -e^{-i2\psi}$ \\[2mm]
 $ (\braketLL{\widehat{p}}{\widecheck{p}}/m)^{1/2}  \equiv  +e^{+i\psi} $ &&
 $(\braketRR{\widecheck{p}}{\widehat{p}}/m)^{1/2} = + e^{-i\psi} $ \\
 \quad $(\braketLL{\widecheck{p}}{\widehat{p}}/m)^{1/2}  = + i e^{+i\psi} $ && \quad
 $  (\braketRR{\widehat{p}}{\widecheck{p}}/m)^{1/2}   =  - i e^{-i\psi} $ \quad\\[2mm]
 \hline
  \end{tabular}
  \caption{ Relationships for the phase factor $\psi$. Note with this set of square root choices, the interchanging $\widehat{p} \leftrightarrow \widecheck{p}$ adds $\pi/2$ to $\psi$ in all of the above and the product of any of phases with its complex conjugate is 1.  }
  \label{tab:phases}
  \end{center}
  \end{table}

The full set of spinors for a massive fermion used in this work is given in Table \ref{tab:Dspinors} and relationships for the phase $\psi$ is given in Table \ref{tab:phases}. All of these spinors satisfy the appropriate Dirac equation,  and are eigenstates of the spin matrix,  $\gamma_5 \slashed{s}$. They are  normalized and orthogonal as follows:
\begin{align}
\overline U_\lambda(p, s) \, U_{\lambda^\prime} (p, s)& = 2m ~\delta_{\lambda {\lambda}^\prime}, \quad \overline V_\lambda(p, s) \, V_{\lambda^\prime} (p, s) = -2m ~\delta_{\lambda {\lambda}^\prime} \,, \notag \\
\overline U_\lambda(p, s) \, V_{\lambda^\prime} (p, s)& = 0 \,,\quad \quad \quad \quad \overline V_\lambda(p, s) \, U_{\lambda^\prime} (p, s) = 0 \,,
 \end{align}
 for all combinations of $\lambda$ and $\lambda^\prime$ equal to  $\uparrow \text{or} \downarrow$. They are connected by

\begin{align}
U_\uparrow(p, s) &=\gamma_5  V_\downarrow(p, s)
\quad \quad U_\downarrow(p, s) =-\gamma_5  V_\uparrow(p, s),\\ 
  \overline V_\uparrow(p, s) & =  
  \overline U_\downarrow(p, s)  \gamma_5, 
 \quad  \;\;  \overline V_\downarrow(p, s)  =   -\overline U_\uparrow(p, s)  \gamma_5 .
  \end{align}
This multiplication by $\gamma_5$ simply changes the sign of one or other of the terms,  since $\gamma_5 = \gamma_R -\gamma_L$. 
Using the charge conjugation matrix, ${\cal C}$, we have
\begin{align}
V_\uparrow (p,s) &=  {\cal C} \biggr(  \overline{U}_\uparrow (p,s) \biggr)^T \,.
\end{align}
With all of these identities this set of spinors are fixed except for an overall phase on
$U_\uparrow, V_\downarrow, \overline{U}_\downarrow, \overline{V}_\uparrow$ and its complex conjugate on
$U_\downarrow, V_\uparrow, \overline{U}_\uparrow, \overline{V}_\downarrow$.\\

For convenience in this paper, we use this phase freedom to connect the following%
\begin{align}   
\overline{U}_\downarrow (p,s) &= - i \, \overline{U}_\uparrow (p,-s),  \quad V_\downarrow (p,s) = - i  \, V_\uparrow (p,-s) \, , 
\label{eq:fixphase}
\end{align}
so that going from $\uparrow$ to $\downarrow$ is simply achieved by interchanging $\widehat{p} \leftrightarrow \widecheck{p}$.
We emphasize that this is just a convenient choice, it is not required by any Quantum Field Theory (QFT) property, and the phase freedom remains, but if used, will break these two conditions. 
The phase choices in ref.~\cite{Mahlon:1995zn} are obtained by multiplying  all of $U_\downarrow, V_\uparrow, \overline{U}_\uparrow, \overline{V}_\downarrow$ by $\exp{+i\psi}$ and all of
$U_\uparrow, V_\downarrow, \overline{U}_\downarrow, \overline{V}_\uparrow$
 by $\exp{-i\psi}$. After this re-phasing the conditions in eq. \eqref{eq:fixphase} do not hold.
 \\

The completeness relationships are
 \begin{align}
 U_{\uparrow} (p, s) \,\overline U_\uparrow(p, s)  = \frac1{2}(1+\gamma_5 \slashed{s})(\slashed{p} +m ), \quad 
& V_{\uparrow} (p, s) \,\overline V_\uparrow(p, s)  = \frac1{2}(1+\gamma_5 \slashed{s})(\slashed{p} - m ) \, , \notag \\[2mm] 
U_{\downarrow} (p, s) \,\overline U_\downarrow(p, s)  = \frac1{2}(1-\gamma_5 \slashed{s})(\slashed{p} +m ), \quad 
& V_{\downarrow} (p, s) \,\overline V_\downarrow(p, s)  = \frac1{2}(1-\gamma_5 \slashed{s})(\slashed{p} - m ) \, , \notag \\[2mm] 
\text{when summed give } \quad \sum_\lambda U_{\lambda} (p, s) \,\overline U_\lambda(p, s)  =\slashed{p} +m  , \quad &\sum_\lambda V_{\lambda} (p, s) \,  \overline V_\lambda(p, s)  = \slashed{p} -m \,.
  \end{align}
We have also explicitly confirmed that the spinors in Table \ref{tab:Dspinors} satisfy all of the identities G.4.34-37 of \cite{Dreiner:2008tw}. \\

With the identities above we can derive relations that all the spinors must satisfy the Dirac equation and also be eigenvectors of the spin matrix, $\gamma_5 \slashed{s}$.
For $U_\uparrow(p, s)$, these are
\begin{align}
 \slashed{p} \, U_\uparrow(p, s) &= m\, U_\uparrow(p, s),  \quad \gamma_5 \slashed{s} \, U_\uparrow(p, s)= + U_\uparrow(p, s) \,.
\end{align}
By adding and subtracting these two identities, after multiplying the second by $m$, we obtain
\begin{align}
[~ \gamma_R \, \slashed{\widehat{p}}+\gamma_L\,\slashed{\widecheck{p}}~]~ U_\uparrow(p, s) & =
[~ \ketR{\widehat{p}}\braL{\widehat{p}} + \ketL{\widecheck{p}}\braR{\widecheck{p}} ~] ~ U_\uparrow(p, s)=  m\, U_\uparrow(p, s) 
, \notag \\ 
[~ \gamma_R\,\slashed{\widecheck{p}}+\gamma_L\,\slashed{\widehat{p}}~]~ U_\uparrow(p, s) & = [~\ketR{\widecheck{p}}\braL{\widecheck{p}} + \ketL{\widehat{p}}\braR{\widehat{p}}~ ] ~ U_\uparrow(p, s) =0 \,.
\end{align}
All the spinors are right or left eigenvectors of the  two matrices 
\begin{align}
    \ketR{\widehat{p}}\braL{\widehat{p}} + \ketL{\widecheck{p}}\braR{\widecheck{p}} ,\quad \ketR{\widecheck{p}}\braL{\widecheck{p}} + \ketL{\widehat{p}}\braR{\widehat{p}}
    \label{eq:projOs}
\end{align}
with eigenvalues, $\pm m$ or $0$, as required, see Table \ref{tab:evalues}. 
All of this is simple to check using spinor-helicity techniques.\\

We remind the reader, that all three polarization vectors for a massive vector particle can be constructed in a similar way using chiral spinors of  $\widehat{p}$ and $\widecheck{p}$, see the appendix of ref.~\cite{Mahlon:1998jd}.\\

\begin{table}
    \begin{tabular}{|c|cccc|cccc|}
    \hline &&&&&&&&\\[-2mm]
    Eigenvalues & ~
    $\overline U_\uparrow(p, s)$~ & $\overline U_\downarrow(p, s) $ ~& $V_\uparrow(p, s)$&$V_\downarrow(p, s)$ ~& ~ $U_\uparrow(p, s)$ ~& $U_\downarrow(p, s)$ ~&
    $\overline V_\uparrow(p, s)$ ~& $\overline V_\downarrow(p, s) $~
    \\[1mm]
    \hline &&&&&&&&\\[-2mm]
    ~$\ketR{\widehat{p}}\braL{\widehat{p}} + \ketL{\widecheck{p}}\braR{\widecheck{p}} $~ &$m$&0&0&$-m$&$m$&0& 0& $-m$ \\[1mm]
    ~$\ketR{\widecheck{p}}\braL{\widecheck{p}} + \ketL{\widehat{p}}\braR{\widehat{p}} $ ~& 0&$m$& $-m$ & 0 & 0&$m$ &$-m$&0 \\[1mm]
    \hline
    \end{tabular} 
  \caption{ The eigenvalues for the two projection matrices given in eq. \ref{eq:projOs}. The sum of the two matrices is just $\slashed{p}$, whereas the difference, line 1 minus line 2, is $m(\gamma_5 \slashed{s})$, demonstrating that each spinor satisfies the appropriate Dirac equation and has the appropriate spin projection. }
  \label{tab:evalues}
\end{table}

\subsection{A Technical Argument Against Anti-symmetrization Option B}
\label{sec:phasechange}

An extra argument can be made against the anti-symmetrization choice of Option B. 
In fact, we can show the interference term for the anti-symmetrization of type B has effectively no physical meaning.
To see this, consider the most general phase change in the spinors given in Table \ref{tab:Dspinors} that is also consistent with QFT
\begin{align}
[U_\downarrow, V_\uparrow, \overline{U}_\uparrow, \overline{V}_\downarrow](p,s)
\rightarrow  & ~
  e^{+i \rho(p,  s)}
[U_\downarrow, V_\uparrow, \overline{U}_\uparrow, \overline{V}_\downarrow](p,s) \,,
 \notag \\
 [U_\uparrow, V_\downarrow, \overline{U}_\downarrow, \overline{V}_\uparrow](p,s)  
\rightarrow & ~
  e^{-i \rho(p,  s)}
  [U_\uparrow, V_\downarrow, \overline{U}_\downarrow, \overline{V}_\uparrow](p,s) 
\,.
  \label{eq:phase_transformation}
\end{align}

The function $\rho$ is quite general, as long as it is  even in the momentum. Using $\Delta \rho \equiv \rho_1 - \rho_2$ and  $\Sigma \rho \equiv \rho_1 + \rho_2$, the change above transforms the Dirac amplitudes by,
\begin{align}
 {\cal D}[\, L_1, \overline{R_2}\, ] \rightarrow
   e^{- i\Delta \rho }~ {\cal D}[\, L_1, \overline{R_2}\, ]
\,, &\quad 
  {\cal D}[\, R_2, \overline{L_1}\, ] 
\rightarrow 
  e^{ -i \Delta \rho } ~{\cal D}[\, R_2, \overline{L_1}\, ]
\,,
\notag \\
{\cal D}[\, L_1, \overline{L_2}\, ] \rightarrow
   e^{- i\Sigma \rho }~ {\cal D}[\, L_1, \overline{L_2}\, ]
\,, &\quad 
  {\cal D}[\, L_2, \overline{L_1}\, ] 
\rightarrow 
  e^{ -i \Sigma \rho } ~{\cal D}[\, L_2, \overline{L_1}\, ] \,,
  \notag \\
  {\cal D}[\, R_1, \overline{R_2}\, ] 
\rightarrow 
  e^{ +i \Sigma \rho } ~{\cal D}[\, R_1, \overline{R_2}\, ] \,,
  & \quad
  {\cal D}[\, R_2, \overline{R_1}\, ] \rightarrow
   e^{+ i\Sigma \rho }~ {\cal D}[\, R_2, \overline{R_1}\, ]
\,, 
  \notag \\
{\cal D}[\, R_1, \overline{L_2}\, ]
\rightarrow 
   e^{+ i \Delta \rho } ~ {\cal D}[\, R_1, \overline{L_2}\, ] \,.
   & \quad    
  {\cal D}[\, L_2, \overline{R_1}\, ]
\rightarrow 
   e^{+ i \Delta \rho } ~ {\cal D}[\, L_2, \overline{R_1}\, ]\,,
\end{align}
Only pairs of amplitudes that have the same phase transformation can be anti-symmetrized without introducing an arbitrary phase between them. These are
\begin{align}
   (L_1\overline{L_2}, ~  \overline{L_1} L_2), \quad 
   (R_1  \overline{R_2}, ~ \overline{R_1} R_2),
   \quad (L_1 \overline{R_2}, ~ \overline{L_1}R_2),
   \quad
   (R_1 \overline{L_2}, ~\overline{R_1}L_2).
\end{align}
These are exactly the pairs that appear in Option A of Fig. \ref{fig:AntiS}.
As an example, consider the $ L_1 \overline{R_2}$ helicity configuration, then the anti-symmetrized amplitude transform as,
\begin{align}
  {\rm Option~A} ~: ~ &
{\cal M}[\, L_1, {R_2}\, ] \equiv {\cal D}[\, L_1, \overline{R_2}\, ] -{\cal D}[\, R_2, \overline{L_1}\, ]  \rightarrow ~ e^{- i \Delta \rho }  \left\{ {\cal D}[\, L_1, \overline{R_2}\, ] -{\cal D}[\, R_2, \overline{L_1}\, ] \right\} \,, \notag
\\ 
  {\rm Option~B} ~: ~ &
~{\cal K}[\, L_1, {R_2}\, ] \equiv {\cal D}[\, L_1, \overline{R_2}\, ] -{\cal D}[\, L_2, \overline{R_1}\, ]  \rightarrow ~  e^{-i \Delta \rho }~{\cal D}[\, L_1, \overline{R_2}\, ] -  e^{+ i \Delta \rho} ~ {\cal D}[\, L_2, \overline{R_1}\, ]\,.  
\end{align}
Clearly, Option A remains invariant, $|{\cal M}[\, L_1, {R_2}\, ]|^2 \rightarrow |
{\cal M}[\, L_1, {R_2}\, ]|^2$, while the interference term of Option B depends on 
the choice of $\Delta \rho$,
\begin{align}
    | {\cal K}[\, L_1, {R_2}\, ]| ^2
\rightarrow 
  |{\cal D}[\, L_1, \overline{R_2}\, ]|^2 
+ |{\cal D}[\, L_2, \overline{R_1}\, ]|^2 
- 2 {\rm Re}\left\{ 
    e^{-2i \Delta \rho}~ {\cal D}[\, L_1, \overline{R_2}\, ]
   ~ {\cal D}^{*}[\, L_2, \overline{R_1}\, ]
  \right\}.
\end{align}
Therefore ${\cal K}[\, L_1, {R_2}\, ]$ cannot be a physical amplitude, further supporting that Option A is the only physics choice.\\

Applying this rephasing to the  sum of the Dirac amplitudes, given by eq. \eqref{eq:KD2},
\begin{align}
&L_1 \overline{R_2} ~+ ~L_1\overline{L_2} ~+~ R_1  \overline{R_2} ~+~ R_1 \overline{L_2} \, 
\quad \Rightarrow  \quad 
e^{- i \Delta \rho }~L_1 \overline{R_2} 
~+ ~e^{- i\Sigma \rho }~L_1\overline{L_2} 
~+~ e^{+ i\Sigma \rho }~R_1  \overline{R_2} 
~+~ e^{+ i\Delta \rho }~R_1 \overline{L_2} 
\end{align}
we see that the relative phase between each amplitude can be varied and therefore this sum cannot be a physical amplitude.

\subsection{Different Bases}
\label{app:bases}

{\bf Charge-Lepton basis:}
In the neutrino rest frame, chose the neutrino spin to be aligned with the direction of one of the charged leptons in that frame and since they are both massless momenta they remain parallel under any boost.\\
For example, choose $\widehat{p}_1 \parallel \bar{k}$ and  $\widehat{p}_2 \parallel k$ then all but one of the Majorana amplitudes in eq.~\eqref{eq:MampS} and \eqref{eq:MampO} are reduced to one term with no interference term. However, one of the amplitudes consists of two terms, in this case ${\cal M}(R,L)$, which leads to the interference term in the amplitude squared summed over spins. \\

{\bf Transverse basis:} in the rest frame of the massive particle the spin axis, say x-axis, is perpendicular to the boost direction, the z-axis, of the particle.
So that before the boost $p=(m,0,0,0)$ 
and $s=(0,1,0,0)$, then after the boost
\begin{align}
p = &  = \gamma m(1, 0,0, \beta), \quad  s=(0,1,0,0), \quad \text{then} \notag \\ 
\widehat{p} = & 
= \frac1{2}\gamma m(1,\gamma^{-1},0,\beta ), 
\quad \widecheck{p} =  \frac1{2}\gamma m(1,-\gamma^{-1},0,\beta ), \notag \\
\text{and} \quad p = & \widehat{p} + \widecheck{p}  \quad \text{with} ~ \widehat{p}^2 = 0=\widecheck{p}^2 \,.
\end{align}
This is easy to generalize to any boost direction and any spin axis perpendicular to the boost.\\

In this basis, the interference terms for the individual Majorana helicity amplitudes, eq. \eqref{eq:Msqed}, is
\begin{align}
{\rm Tr}[k \widecheck{p}_1 \widehat{p}_1
\bar{k} \widecheck{p}_2 \widehat{p}_2]/m^2 \leq 4 \sqrt{(k \cdot \widecheck{p}_1)(\widehat{p}_1 \cdot
\bar{k}) (\bar{k} \cdot \widecheck{p}_2) (\widehat{p}_2 \cdot k)}  \sim {\cal O}( \, E_1 E_2 |k| |\bar{k}|\,) 
\end{align}
and therefore is not suppressed by the neutrino mass.  However, due to cancellations, the sum of the interference terms is still suppressed by the neutrino mass, as this sum is $m_\nu^2 (k\cdot \bar{k})$ as given in eq. \eqref{eq:totM}.  This is related to the decoupling of the transverse degrees of freedom as the fermion mass goes to zero, which was further discussed in Ma and Pantaleone~\cite{Ma:1989jpa}, see Appendix~\ref{app:MP}. \\

\section{\texorpdfstring{Application to $e^+e^- \rightarrow NN$}{Application to eeNN}}
\label{app:MP}
The fermionic component of the leptonic $\Phi$ decay in Section~\ref{sec:Scalar} may be related to the process $e^+ e^- \to NN$ by a crossing symmetry, where the $N$ are heavy Majorana fermions with mass $m_N$. Therefore, the spinorial structure of the result for this process by Ma and Pantaleone~\cite{Ma:1989jpa} can be compared with that calculated here as an independent check of our result.  Expanding, for example, from our $RR$ Majorana amplitude squared, we obtain
\begin{align}
|\, {\cal M}[\, R_1, {R_2}\, ]  \,|^2 &=16(k \cdot \widecheck{p}_1)(\bar{k}\cdot \widehat{p}_2)
+16(k \cdot \widecheck{p}_2)(\bar{k}\cdot \widehat{p}_1) -4\,\text{Tr}[k \widecheck{p}_1 \widehat{p}_1
\bar{k} \widehat{p}_2 \widecheck{p}_2]/m_N^2 \label{eq:RRresult} \\
=4m_N&[ \bar{k}\cdot (s_2-s_1)\,k\cdot (p_1-p_2)- k\cdot (s_2-s_1)\bar{k}\cdot (p_1-p_2)]\notag\\
& +4 (\bar{k}\cdot p_2)(k\cdot p_1) +4(\bar{k}\cdot p_1) (k\cdot p_2)  -4m_N^2(k\cdot\bar{k})+4(p_1\cdot p_2-m_N^2)\left[(  s_2 \cdot \bar{k})(s_1\cdot k)+(s_2\cdot k)(s_1\cdot \bar{k})\right]\notag\\
&-4(s_2\cdot p_1)\left[ (p_2\cdot \bar{k})(s_1\cdot k) +(p_2\cdot k)(s_1\cdot \bar{k})\right]+4(s_2\cdot p_1)(\bar{k}\cdot k )(p_2\cdot s_1)-4(s_2\cdot s_1)(\bar{k}\cdot k)( p_2\cdot p_1) \notag\\
&+4(s_2\cdot s_1)\left[(\bar{k}\cdot p_2)(k\cdot p_1)+(k\cdot p_2)(\bar{k}\cdot p_1) \right] -4(p_2\cdot s_1) \left[(s_2\cdot \bar{k})(p_1\cdot k)+(s_2\cdot k)(p_1\cdot  \bar{k})\right] \label{eq:eeNN}\end{align}
This may be used to calculate the left-chiral current contribution to $e^+e^-\to NN$ by performing the substitution $\bar{k}\to -{k}$ and ${k}\to -\bar{k}$, where we have insisted on continuity that the ``barred" momentum represents the charged anti-fermion momentum. Then it may be contrasted with eq. (1) of ref.~\cite{Ma:1989jpa}, with $x=\sin^2\theta_W\mapsto 0$, with which we find perfect agreement for each of the spin contributions.\footnote{The terms are organized above for easy contrast with eq.(1) of Ma and Pantaleone~\cite{Ma:1989jpa}. After performing the $\bar{k}\to -{k}$ and ${k}\to -\bar{k}$ substitution, matching onto the notation of \cite{Ma:1989jpa} defined in their Figure 1 involves the following further notation changes: $k\mapsto p, \bar{k}\mapsto \bar{p}, s_1\mapsto s, s_2\mapsto \bar{s}, p_1\mapsto q$ and $p_2\mapsto \bar{q}$. } This demonstrates the power of the spinor-helicity formalism as applied to these processes: the result in eq.~\eqref{eq:RRresult} is more compact at illuminating the underlying physics than that from traditional trace techniques. \\

The result in eq.~\eqref{eq:eeNN} contains interference terms that scale with $m_N^2$, and others which represent transverse degrees of freedom, as discussed in ref.~\cite{Ma:1989jpa}. Summing over spins for the result in eq.~(1) of ref.~\cite{Ma:1989jpa} of  is equivalent to summing over helicity configurations for the massless spinors constructed with $\widehat{p}, \widecheck{p}$. Doing so reveals that these transverse terms will cancel, and we recover interference behavior in the total squared matrix-element consistent with that described in eq.~\eqref{eq:totM}.  To obtain the additional contributions to $e^+e^-\to NN$ in ref.~\cite{Ma:1989jpa} (\emph{i.e.} where $x\neq 0$) one needs to consider different chiralities of interactions which may be generated by $Z$ exchange, which may straightforwardly be derived using the same spinor-helicity methods as performed in this work.


\acknowledgments

This manuscript has been authored in part by FermiForward Discovery Group, LLC under Contract Number 89243024CSC000002 with the U.S. Department of Energy, Office of Science, Office of High Energy Physics. IB was supported by the Munich Institute for Astro-, Particle and BioPhysics (MIAPbP) which is funded by the Deutsche Forschungsgemeinschaft (DFG, German Research Foundation) under Germany's Excellence Strategy – EXC-2094 – 390783311. PP is supported by FAEPEX grant number 2404/25 FAEPEX/UNICAMP. This project has also received support from the European Union's Horizon 2020 research and innovation programme under the Marie  Sklodowska-Curie grant agreement No 860881-HIDDeN as well as under the Marie Skłodowska-Curie Staff Exchange grant agreement No 101086085 - ASYMMETRY. IB and SP acknowledge discussions with C.S.~Kim at Fermilab that evolved our understanding of refs.~\cite{Kim:2021dyj, Kim:2023iwz}, leading to Sec. \ref{sec:b0decay}.

\newpage
\bibliographystyle{utphys}
\bibliography{main_bibl}
\end{document}